\documentclass{emulateapj}

\newcommand{\chandra}{{\it Chandra}}
\newcommand{\swift}{{\it Swift}}
\newcommand{\xmm}{{\it XMM-Newton}}
\newcommand{\suzaku}{{\it Suzaku}} 

\begin{document}

\markboth{J.-U. Ness et al.}{\xmm\ X-ray and UV observations of V2491\,Cyg}

\title{\xmm\ X-ray and UV observations of the fast nova V2491\,Cyg during the supersoft source phase}
\author{J.-U. Ness\altaffilmark{1},
  J.P. Osborne\altaffilmark{2},
  A. Dobrotka\altaffilmark{3,4},
  K.L. Page\altaffilmark{2},
  J.J. Drake\altaffilmark{5},
  C. Pinto\altaffilmark{6},
  R.G. Detmers\altaffilmark{6},
  G. Schwarz\altaffilmark{7},
  M.F. Bode\altaffilmark{8},
  A.P. Beardmore\altaffilmark{2}
  S. Starrfield\altaffilmark{9},
  M. Hernanz\altaffilmark{10},
  G. Sala\altaffilmark{11},
  J. Krautter\altaffilmark{12},
  C.E. Woodward\altaffilmark{13}
}

\altaffiltext{1}{XMM-Newton Science Operations Centre, ESA, PO Box 78, 28691 Villanueva de la Ca\~nada, Madrid, Spain}
\altaffiltext{2}{Department of Physics \& Astronomy, University of
  Leicester, Leicester, LE1 7RH, UK}
\altaffiltext{3}{Department of Physics, Institute of Materials Science, Faculty of Materials Science and Technology, Slovak University of Technology in Bratislava, J\'ana Bottu 25, 91724 Trnava, The Slovak Republic}
\altaffiltext{4}{Department of Astronomy, Graduate School of Science, Kyoto University, Sakyo-ku, Kyoto 606-8502, Japan}
\altaffiltext{5}{Harvard-Smithsonian Center for Astrophysics, 60
  Garden Street, Cambridge, MA 02138, USA}
\altaffiltext{6}{SRON Netherlands Institute for Space Research,
 Sorbonnelaan 2, 3584 CA Utrecht, the Netherlands}
\altaffiltext{7}{American Astronomical Society, 2000
   Florida Ave., NW, Suite 400, Washington, DC 20009-1231, USA}
\altaffiltext{8}{Astrophysics Research Institute, Liverpool John
  Moores University, Twelve Quays House, Egerton Wharf, Birkenhead
  CH41 1LD, UK}
\altaffiltext{9}{School of Earth and Space Exploration, Arizona
State University, Tempe, AZ 85287-1404, USA}
\altaffiltext{10}{Institut de Ci\`encies de l'Espai (CSIC-IEEC),
Campus UAB, Facultat de Ci\`encies, C5 parell 2$^{on}$,
08193 Bellaterra (Barcelona), Spain}
\altaffiltext{11}{Departament F\'{\i}sica i Enginyeria Nuclear, EUETIB (UPC-IEEC),
Comte d'Urgell 187, 08036 Barcelona, Spain}
\altaffiltext{12}{Landessternwarte K\"onigstuhl, 69117 Heidelberg, Germany}
\altaffiltext{13}{Department of Astronomy, University of Minnesota, 116 Church Street
SE, Minneapolis, MN 55455, USA}

\begin{abstract}
 Two \xmm\ observations of the fast classical nova V2491\,Cyg were
carried out in short succession on days 39.93 and 49.62 after discovery,
during the supersoft source (SSS) phase, yielding simultaneous X-ray and UV
light curves and high-resolution X-ray spectra. The first X-ray light
curve is highly variable, showing oscillations with a period of 37.2
minutes after an extended factor of three decline lasting $\sim$ 3 hours,
while the second X-ray light curve is less variable.
The cause of the dip is currently unexplained and
has most likely the same origin as similar events in the early SSS
light curves of the novae
V4743\,Sgr and RS\,Oph, as it occurred on the same time scale.
The oscillations are not present during
the dip minimum and also not in the second observation.
The UV light curves are variable but contain no dips and no period.
High-resolution X-ray spectra are presented for 4 intervals of
differing intensity. All spectra are atmospheric
continua with deep absorption lines and absorption edges. Two
interstellar lines of O\,{\sc i} and N\,{\sc i} are clearly seen
at their rest wavelengths, while a large number of high-ionization
absorption lines are found at blue shifts indicating an expansion
velocity of $3000-3400$\,km\,s$^{-1}$, which does not change
significantly during the epochs of observation. Comparisons with the
slower nova
V4743\,Sgr and the symbiotic recurrent nova RS\,Oph are presented.
The SSS spectrum of V4743\,Sgr is much softer with 
broader and more complex photospheric absorption lines. The ejecta are
extended, allowing us to view a larger range of the
radial velocity profile. Meanwhile, the absorption lines in
RS\,Oph are as narrow as in V2491\,Cyg, but they are less
blue shifted. A remarkable similarity in the continua of V2491\,Cyg
and RS\,Oph is found. The only differences are
smaller line shifts and additional emission lines in RS\,Oph
that are related to the presence of a dense stellar wind from the
evolved companion. 
Three unidentified absorption lines are present in the X-ray spectra of
all three novae, with projected rest wavelengths
26.05\,\AA, 29.45\,\AA, and 30.0\,\AA.
No entirely satisfactory spectral model is currently available for the
soft X-ray spectra of novae in outburst, and careful discussion of
assumptions is required.
\end{abstract}


\section{Introduction}

Classical Nova (CN) and Recurrent Nova (RN) outbursts result from nuclear
explosions on the surface of a white dwarf (WD) that has accreted
hydrogen-rich material from a companion star, which in the case of
most CNe is a low mass main sequence star. Some RNe occur in
symbiotic systems where the companion is evolved.
Enough energy is produced to eject the
outer envelope. The radiative output of the nova first
appears primarily in the optical, but as the density in the
ejecta drops as a consequence of decreasing mass loss rate from the
WD surface and continued expansion, the radius of
the pseudo-photosphere shrinks, and successively hotter layers
become visible \cite[see, e.g.,][]{gallstar78,hauschildt92}. If
nuclear burning continues long enough, the peak of the spectral energy
distribution eventually shifts into the X-ray regime, and at that time
the nova emits an X-ray spectrum that resembles those typically
observed in the class of supersoft X-ray sources
\citep[SSS:][]{kahab}. For a review of nova evolution see, e.g.,
\cite{bodeevansbook}.
 
 In the standard picture of nova evolution, the only time at which
novae are bright X-ray emitters is during this
SSS phase, but it is difficult to predict at which time this phase
starts. In an attempt to catch a nova during the SSS phase, early
observations have to be taken that bear the risk of no detection.
During these campaigns, several novae have been observed before the
SSS phase started, yielding faint, hard X-ray spectra
\citep[e.g.,][]{lloyd92,krautt96,mukai01,Orio2001}. The
early faint, hard X-ray emission is believed to arise from shocks
in the ejecta or in specific cases such as RS\,Oph and other
symbiotic systems, in shocks between the ejecta and the dense
stellar wind of the companion. In very special cases such as
V458\,Vul, early hard emission may also arise from interactions
of the ejecta with surrounding material in a pre-existing
planetary nebula \citep[][and references therein]{Wessonetal08,v458}.

The uncertainty of the times at which bright SSS emission can be
seen has been significantly reduced by systematic monitoring with the
X-ray Telescope (XRT) on board the \swift\ observatory. The first
dense \swift\ monitoring
campaign was carried out after the 2006 outburst of RS\,Oph
\citep{bode06,osborne11,page08}, and
\xmm\ and \chandra\ observations were scheduled at times that were
strategically important \citep[e.g.][]{ness_rsoph,rsophshock}.

 V2491 Cyg was discovered on 2008 April 10.728UT by
\cite{v2491discovery}. The optical brightness was 7.7 mag at the
time of discovery and it reached a peak magnitude of 7.5 mag one
day later. The time scale by which the V brightness decreased by 
two magnitudes was $t_2=4.6$ days \citep{tomov08b}), as such a
very fast nova. A rebrightening has been observed after
JD$\sim 2454575$ (April 26). \cite{naik09} discussed this
secondary peak. In early spectroscopic
observations, P-Cygni line profiles were found with a velocity of
about $-4000$\,km\,s$^{-1}$ in H$\alpha$ and H$\beta$ profiles
\citep{tomov08a}.
Near infrared (NIR) photometry \citep{naik09} and spectroscopy
\citep{rudy08}
obtained in April, 2008 showed no sign of dust formation.
Unpublished NIR spectroscopy taken in July 2008 also revealed no
dust formation (Rudy, private communication), indicating that
V2491\,Cyg was not a dust forming nova.
\cite{rudy08} measured $E(B-V)=0.43$, which converts to an
interstellar neutral hydrogen column density of
$N_{\rm H}=2.6\times10^{21}$\,cm$^{-2}$ using the relation
$<N_{\rm H}$/E(B-V)$>=6\,\pm\,2\times10^{21}$\,cm$^{-2}$
\citep{dickey90,bohlin78}.

 The distance was estimated to be 10.5\,kpc by \cite{helton08}
and, more recently, 14\,kpc by \cite{munari10}, both based on a
Maximum Magnitude versus Rate of Decline relation by \cite{DVL95,downes00}.
We note, however, that in addition to the high uncertainty of this
method \citep[see, e.g.,][]{warner08}, a second, smaller, maximum
occurred about three weeks after discovery, which reduces the
confidence of this method. This makes the time zero from which
the MMRD relation should be used uncertain.

 In X-rays, V2491\,Cyg was observed with \swift\ 1.02 days after
discovery, but only a marginal detection was achieved
\citep{page09}, which could, however, be contaminated by optical
loading, i.e., the high number of optical photons leads to
false registration of X-ray events
\citep{v2491prenova_atel}. During a second
\swift\ observation on day 4.56 after discovery, a count rate of
$0.009\,\pm\,0.002$ counts per second (cps) was reported by
\cite{v2491firstswift}. In a \suzaku\ observation taken 9 days after
discovery, superhard X-ray emission extending out to 70\,keV was
detected in addition to a hard thermal spectrum \citep{v2491_superhard}.

This nova is only the second nova to be detected in X-rays before
the outburst, after V2487 Oph \citep{v2487_prenova}.
\cite{v2491prenova} reported on five \swift\ pre-outburst
observations in which V2491\,Cyg was serendipitously in the field
of view of the XRT with clear detections.

 Based on the optical spectral characteristics of V2491 Cyg, \cite{tomov08b}
speculated that it is a recurrent nova, similar to U Sco and V394 CrA.
The similarity to V2487\,Oph with regards to pre-outburst X-ray
emission and the identification of V2487\,Oph as a recurrent nova
by \cite{v2487oph_rn} (they discovered a prior outburst in 1900)
support this speculation.

The evolution of this nova was followed with \swift\
observations every 1-3 days. \cite{page09} give
a detailed description of the evolution of X-ray brightness,
X-ray spectra, and UV brightness.

\begin{figure}
\resizebox{\hsize}{!}{\includegraphics{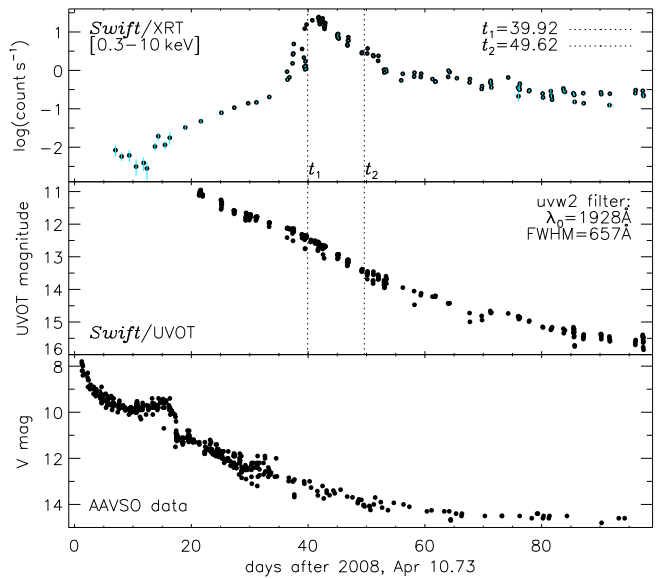}}
\caption{\label{swlc}\swift\ X-ray (top) and UV (middle) light
curves from Page et al. (2010) with times of \xmm\ observation
marked by vertical dotted lines. In the bottom panel, the AAVSO
light curve is added, showing the evolution in the V filter. A
small drop in X-ray count rate a few days before the secondary
peak can be seen.
}
\end{figure}

 The \swift\ monitoring observations were used to schedule two
\xmm\ observations as illustrated in Fig.~\ref{swlc}
\citep{v2491_xmm1,v2491_xmm2}, where, from top to bottom,
light curves are shown from \swift/XRT and UVOT (data taken from
\citealt{page09}), and V filter magnitudes from the
database of the American Association of Variable Star
Observers (AAVSO).
The secondary peak can be seen in the AAVSO light curve around
day 15, a few days after a small reduction in the X-ray brightness.

 The first \xmm\ observation was carried out five
days after the first detection of SSS emission. As soon
as the indication of a decline was found, a second \xmm\ observation
was obtained. A rich dataset was acquired, including X-ray and
UV light curves, high-resolution X-ray spectra in the energy range
0.3-2\,keV, and low-resolution X-ray spectra in the range 0.1-10\,keV.
In this paper, the soft X-ray spectra and X-ray and UV light curves
are presented and described.
In \S\ref{obssect} we describe the observations and the data reduction.
The light curves are presented and analyzed in \S\ref{lcsect}, and
the presentation of X-ray spectra can be found in \S\ref{spectra}.
These descriptions need to be complemented
by models that will be developed and discussed by theoretical
working groups, and are therefore not discussed beyond best fits of
publicly available models in \S\ref{atmsect}.
The results from light curve and spectral analyses are discussed in
\S\ref{disc} with a focus on spectral modeling, comparison with
X-ray spectra of other novae, and high-amplitude variations.
Summary and conclusions are given in \S\ref{summary}.

\section{Observations}
\label{obssect}

 Two \xmm\ observations of V2491\,Cyg were taken, starting on
2008 May 20.6 (39.93 days after discovery) and May 30.3 (49.62 days
after discovery). The \xmm\ observatory consists of five different
instruments behind three mirrors plus an optical monitor that all
observe simultaneously. The observation details are listed
in Table~\ref{tab1}.

 For this paper, only the spectra from the Reflection Grating
Spectrometers (RGS, \citealt{rgs}), the light curves from the
EPIC/pn, and the optical monitor (OM, \citealt{om}) are used.
As will be shown, the SSS spectra range from $15-38$\,\AA\
(0.33-0.83\,keV), a range completely covered by the RGS. Since the
RGS spectrum is sufficiently well exposed for our analysis,
the EPIC spectra of this spectral range are not needed.
At energies above 2\,keV, optically thin
thermal emission is observable in the EPIC spectra. These are
discussed in the context of additional Suzaku observations by
\cite{dai_v2491}.

The RGS consists of two identical grating spectrometers, RGS1 and
RGS2, behind different mirrors. The dispersed photons are
recorded by a strip of eight CCD Metal Oxide Semi-conductor
(MOS) chips. One of these chips has failed in each spectrometer
leading to gaps in the spectra, fortunately affecting different
spectral regions, and the missing information can be retrieved
from the respective other spectrometer. A number
of spectral bins contain no information because of bad pixels.
Again, the redundancy of having two spectrometers allows almost
all gaps to be filled.

 The OM was operated in Science User
Defined imaging plus fast mode, which yields UV light curves. Several
shorter exposures were taken with the UVW1 ($\lambda\sim 2500-3500$\,\AA),
UVM2 ($\lambda\sim 2000-2600$\,\AA), and UVW2 ($\lambda\sim 1900-2300$\,\AA)
filters.
Some of the OM exposures failed because of double bit memory errors
(listed with 0 exposure time in Table~\ref{tab1}).
These exposures are not recoverable.

\begin{figure}
\resizebox{\hsize}{!}{\includegraphics{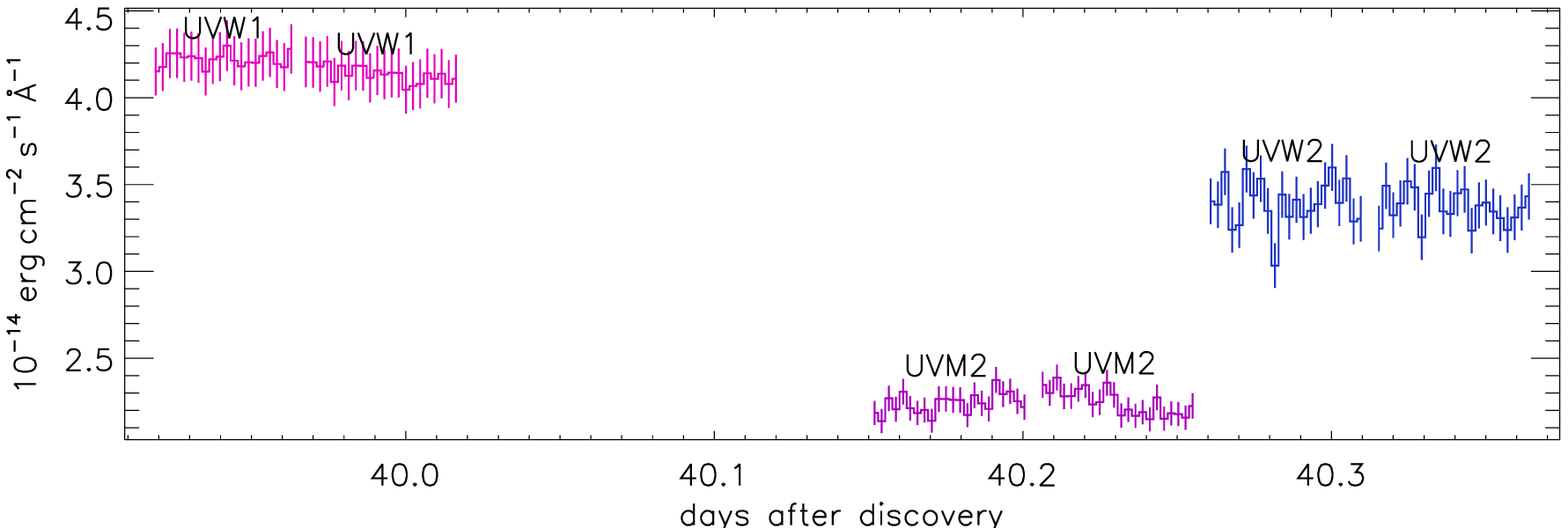}}
 
\resizebox{\hsize}{!}{\includegraphics{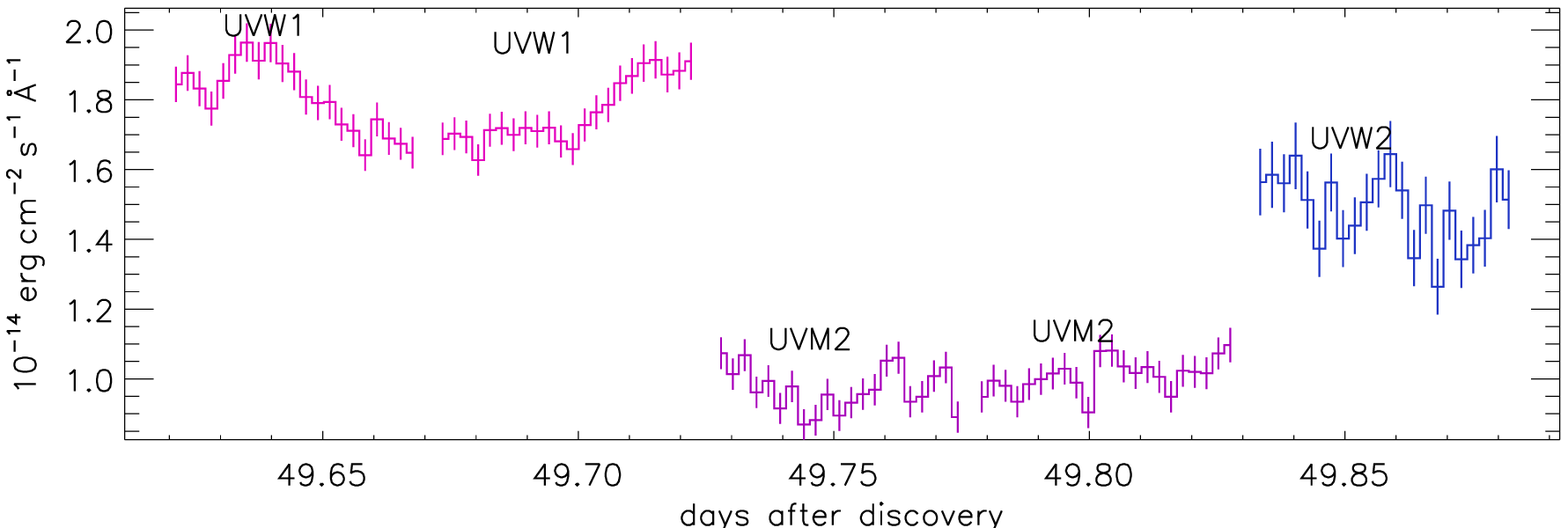}}  
\caption{\label{lcuv}OM light curves in flux units.
 The time axis is plotted in units of fractional day
after discovery (April 10.728, 2008).
The time range between days 40.02 and 41.15 are not
covered because of double bit memory problems leading
to unrecoverable loss of data.
}
\end{figure}

In Table~\ref{uv}, the average magnitudes and fluxes in each
filter are listed for the exposures taken at the times given in
the first column. The UVW2 magnitudes are consistent with the
\swift/UVOT measurements reported by \cite{page09} for the
corresponding times. A graphical illustration of OM fluxes
is shown in Fig.~\ref{lcuv}.

 The EPIC/pn was operated in timing mode. The events are all collapsed
into a single column, allowing fast readout of the chips. While the
high time resolution is not critical to our analysis, the faster
read out prevents pile up and the extracted spectra and light curves
can be used without non-standard corrections.

\begin{table*}
\begin{flushleft}
\caption{\label{tab1}Observation log}
\begin{tabular}{lllllll}
\hline
Instrument & Mode  & Filter & Start Time & End Time & Scheduled & Performed \\
\multicolumn{3}{l}{\bf ObsID 0552270501}&\multicolumn{2}{c}{YYYY-MM-DD@HH:MM:SS}&Duration (sec) &Duration (sec)\\
MOS1 & Small Window & MEDIUM & 2008-05-20@14:04:42 & 2008-05-21@00:58:39 & 39057 & 39057\\
MOS2  & Small Window  & MEDIUM & 2008-05-20@14:04:42  & 2008-05-21@00:58:44  & 39062  & 39062\\ 
pn  & Timing  & MEDIUM & 2008-05-20@14:24:03  & 2008-05-21@00:58:59  & 38036  & 38036\\ 
RGS1  & \multicolumn{2}{l}{Spectroscopy SES} & 2008-05-20@14:03:29  & 2008-05-21@00:59:54  & 39283  & 39283\\
RGS2  & \multicolumn{2}{l}{Spectroscopy SES} & 2008-05-20@14:03:34  & 2008-05-21@00:59:54  & 39278  & 39278\\
OM  &     & UVW1  & 2008-05-20@14:09:10  & 2008-05-20@15:19:16  & 3900  & 3900\\
OM  &     & UVW1  & 2008-05-20@15:19:17  & 2008-05-20@17:07:43  & 4400  & 4400\\OM  &     & UVW1  & 2008-05-20@17:07:44  & 2008-05-20@18:26:10  & 4400  & 0$^a$\\
OM  &     & UVM2  & 2008-05-20@18:26:11  & 2008-05-20@19:44:37  & 4400  & 0$^a$\\
OM  &     & UVM2  & 2008-05-20@19:44:38  & 2008-05-20@21:03:04  & 4400  & 4400\\
OM  &     & UVM2  & 2008-05-20@21:03:05  & 2008-05-20@22:21:31  & 4400  & 4400\\
OM  &     & UVW2  & 2008-05-20@22:21:32  & 2008-05-20@23:39:58  & 4400  & 4400\\
OM  &     & UVW2  & 2008-05-20@23:39:59  & 2008-05-21@00:58:25  & 4400  & 4400\\
\hline
\multicolumn{7}{l}{\bf ObsID 0552270601}\\
MOS1 & Small Window & THIN 1 & 2008-05-30@08:21:53 & 2008-05-30@16:39:10 & 29657 & 31150\\
MOS2  & Small Window  & THIN 1  & 2008-05-30@08:21:53  & 2008-05-30@16:39:15  & 29662  & 31165\\
pn  & Timing  & MEDIUM & 2008-05-30@08:41:14  & 2008-05-30@16:39:30  & 28636  & 30208\\ 
RGS1  & \multicolumn{2}{l}{Spectroscopy SES} & 2008-05-30@08:20:40  & 2008-05-30@16:40:25  & 29883  & 31725\\
RGS2  & \multicolumn{2}{l}{Spectroscopy SES} & 2008-05-30@08:20:45  & 2008-05-30@16:40:25  & 29878  & 31657\\
OM  &    & UVW1  & 2008-05-30@08:26:21  & 2008-05-30@09:41:27  & 4200  & 4200\\
OM  &    & UVW1  & 2008-05-30@09:41:28  & 2008-05-30@10:59:54  & 4400  & 4400\\
OM  &    & UVM2  & 2008-05-30@10:59:55  & 2008-05-30@12:13:21  & 4100  & 4100\\
OM  &    & UVM2  & 2008-05-30@12:13:22  & 2008-05-30@13:31:48  & 4400  & 4400\\
OM  &    & UVW2  & 2008-05-30@13:31:49  & 2008-05-30@14:50:15  & 4400  & 4400\\
OM  &    & UVW2  & 2008-05-30@14:50:16  & 2008-05-30@16:38:42  & 4400  & 0$^a$\\
\hline
\end{tabular}

$^a$double bit memory errors
\renewcommand{\arraystretch}{1}
\end{flushleft}
\end{table*}

\begin{figure}
\resizebox{\hsize}{!}{\includegraphics{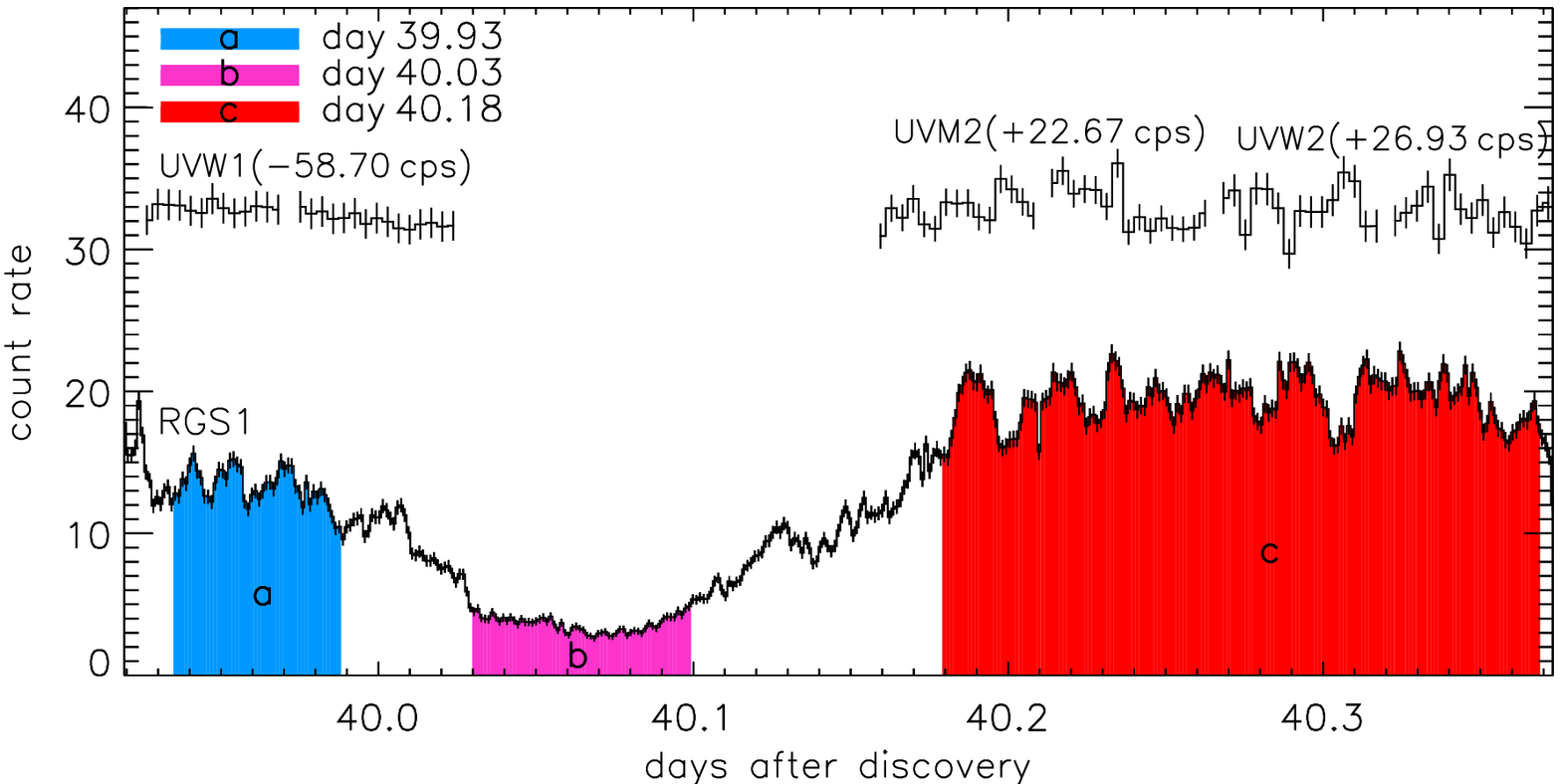}}

\resizebox{\hsize}{!}{\includegraphics{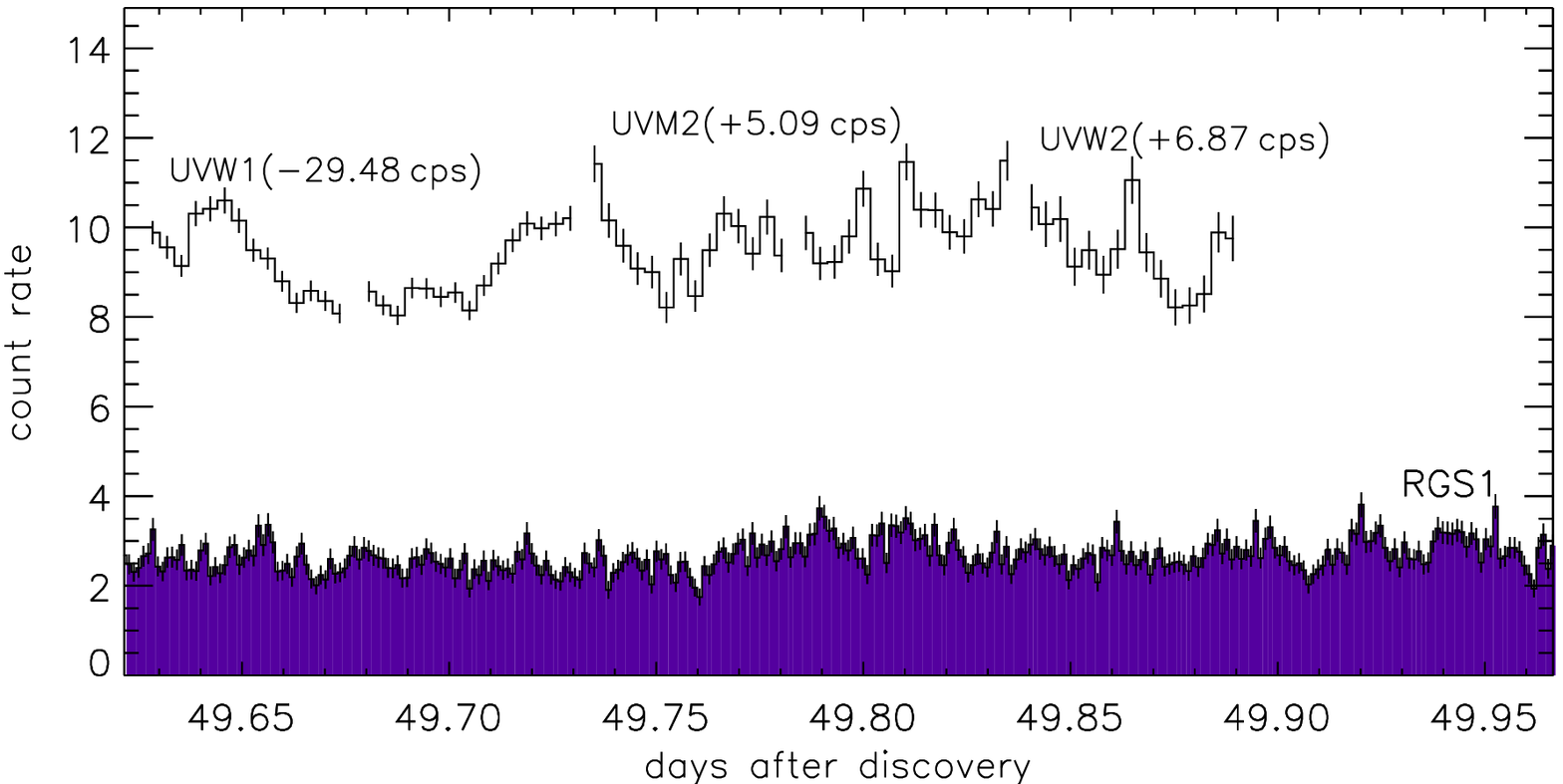}}

\caption{\label{lc}X-ray RGS1 and OM light curves for the observations
starting on days 39.93 (top) and 49.62 (bottom). Time bin sizes are
50 seconds and 300 seconds for RGS1 and OM, respectively. The OM count
rates were rescaled to fit in the graph. The shaded areas
mark time intervals over which we have extracted separate RGS spectra
for studies of spectral changes with the photometric variability
(see Table~\ref{sepoch}).
}
\end{figure}

 We have used standard SAS (Science Analsis Software, version 10.0) tools
for the reduction of light curves and spectra. The new SAS tool {\tt
xmmextractor} was particularly helpful in determining optimized extraction
regions and the correction of telemetry losses in the EPIC/pn light
curve.\footnote{Many thanks to Aitor Ibarra for support with {\tt xmmextractor}.}

\section{Light Curves}
\label{lcsect}

In Fig.~\ref{lc} we present the X-ray light curves extracted from the
RGS1 in comparison with the simultaneously taken ultraviolet light
curves shown in Fig.~\ref{lcuv}. The UV fluxes are shifted and
rescaled to fit in the sample plot. The time axis is plotted in
units of fractional day after discovery (April 10.728, 2008).

\begin{table}
\begin{flushleft}
\caption{\label{uv}Ultraviolet magnitudes and fluxes for three OM filters}
\begin{tabular}{lp{.4cm}p{1.2cm}p{.4cm}p{1.cm}p{.4cm}p{1.5cm}}
\hline
& \multicolumn{2}{c}{UVW1}&
 \multicolumn{2}{c}{UVM2}&
\multicolumn{2}{c}{UVW2}\\
& \multicolumn{2}{c}{$\lambda_{\rm eff}^a=2910$\,\AA}&
 \multicolumn{2}{c}{$\lambda_{\rm eff}^a=2310$\,\AA}&
\multicolumn{2}{c}{$\lambda_{\rm eff}^a=2120$\,\AA}\\
Day & mag & \hfill flux$^b$& mag & \hfill flux$^b$& mag & \hfill flux$^b$\\
\hline
39.93 & 12.29& \mbox{$4.39\,\pm\,0.005$} &&&&\\
39.98 & 12.30& \mbox{$4.36\,\pm\,0.013$} &&&&\\
40.16 &&& 13.17&\mbox{$2.42\,\pm\,0.01$} &&\\
40.21 &&& 13.17&\mbox{$2.42\,\pm\,0.01$} &&\\
40.27 &&&&& 12.89&\mbox{$3.53\,\pm\,0.021$}\\
40.32 &&&&& 12.90&\mbox{$3.50\,\pm\,0.021$}\\
49.63 & 13.21& \mbox{$1.89\,\pm\,0.004$} &&&&\\
49.68 & 13.23& \mbox{$1.85\,\pm\,0.004$} &&&&\\
49.73 &&& 14.06&\mbox{$1.06\,\pm\,0.013$} &&\\
49.78 &&& 14.03&\mbox{$1.10\,\pm\,0.012$} &&\\
49.84 &&&&& 13.79&\mbox{$1.54\,\pm\,0.024$}\\
\hline
\end{tabular}

$^a$effective wavelength\\
$^b 10^{-14}$\,erg\,cm$^{-2}$\,s$^{-1}$\AA$^{-1}$
\end{flushleft}
\end{table}

In the first observation, starting 39.93 days after discovery, the X-ray
light curve is highly variable, while the UV light curves appear less
variable. We have checked for variability in the light curves in both bands.
For each OM exposure, we tested the original (non-scaled) count rate,
to be constant at the median value and calculated a value of reduced
$\chi_\nu^2$. All exposures yield $\chi_\nu^2\sim 1$,
indicating no significant variability longer than the binsize of 50 seconds.
 The most prominent event in the X-ray light curve is an extended
dip in count rate, commencing on day 40.01, lasting 0.12 days (2.9 hours).
After this event, the X-ray count rate rises again, yielding a higher count
rate than before the dip. Unfortunately, two OM exposures failed during
the most interesting time interval between 40.02 and 40.15 days. We are
therefore unable to study any relations between the dip in X-rays and the UV.
In addition to the dip, shorter variations of order 15 min can be seen
in the X-ray count rate before day 40.0 and after day 40.12 which could
be oscillations (see \S\ref{lcmod} below). We fitted a 5th order polynomial
to the entire RGS1 X-ray light curve. Although the polynomial describes
the long-term trend well, we only found a best-fit value of
$\chi^2_\nu=10.3$ (with 781 degrees of freedom), thus additional
variability on shorter time scales is present at a significant level.
 Interestingly, during the dip, no oscillations seem to be
present which we verified by computing $\chi^2_\nu$ for the time interval
during the dip, finding $\chi^2_\nu<1$ after fitting a 5th-order
polynomial to this subset as well as for a constant count rate.
For the part after the dip, we find $\chi^2_\nu=7.2$ (340 degrees of
freedom) when compared to a 5th order polynomial,
a bit better, but indicating that there is more variability in addition to
the dip, motivating a detailed analysis of variability in \S\ref{lcmod}.

 The shaded regions indicate three time intervals for which we have
extracted RGS spectra that are discussed in \S\ref{spectra}. We refer to
these episodes as phases a, b, and c as indicated inside the shaded
areas and the legend (see also Table~\ref{sepoch}). Phase a is the
pre-dip spectrum, phase b corresponds to the time of low-flux emission,
and phase c is the post-dip.

While the OM light curve appears to show some anticorrelated
variations to the X-ray light curves, these cannot be considered
significant as the OM light curve is not variable on a statistically
significant level. When ignoring the measurement errors, a Spearman
Rank test yields no correlation or anticorrelation between the X-ray
and UV light curves.

During the second observation, the X-ray brightness (binned in 50
second bins) deviates only marginally from a constant rate, yielding
$\chi^2_\nu\sim 2.9$ with 594 degrees of freedom. For the OM light
curves, binned on 50-second grids, the assumption of constant light
curves yields $\chi^2_\nu$ of 1.2 for the
UVW2 light curve (which has the lowest average count rate),
1.6 for the two UVM2 light curves, but 4.2 and 4.6 for
the best-exposed UVW1 light curves. All fits have been done with
42 degrees of freedom.

 The higher degree of variability in X-rays during the first
observation, especially with the deep dip, is consistent with a high
degree of variations in the early \swift/XRT light curve
(see Fig.~\ref{swlc}), while the second observation was taken at a
time when the \swift\ X-ray light curve was also more stable with no
dips \citep{page09}.
Meanwhile, the long-term UV light curve taken with the UVOT on board
\swift, shows a slow, continuous decay with no signs of variability
(see middle panel of Fig.~\ref{swlc}).
The \xmm\ light curves fit into this general picture with a high
degree of variability in X-rays and a lower degree of variability in
the UV.

\subsection{Timing analysis}
\label{lcmod}

 We have performed period studies using the EPIC/pn, EPIC/MOS, and
RGS light curves of the first observation, all giving consistent
results. As discussed above, the OM light curves and the second
observation are not significantly variable for period studies.
In order to remove
long-term trends, we have detrended the light curves with an
8th-order polynomial and applied a Lomb-Scargle algorithm
\citep{scargle82} and a 2-sine fitting method developed by
\cite{dobrness09}. 99.7\% errors are calculated from the false
alarm probability \citep{hornebal} for the Lomb-Scargle analysis
and from 3-$\sigma$ confidence isocontours for the 2-sine
fitting method. Since no variability is present during the
dip, we have excluded the data between 8 and 22 ks of elapsed
time.

\begin{figure}
\resizebox{\hsize}{!}{\includegraphics{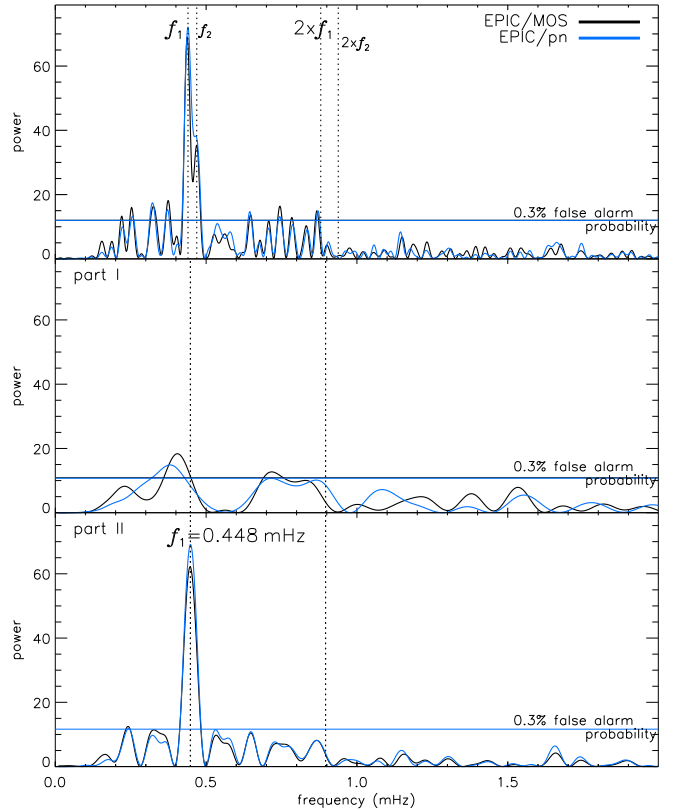}}
\caption{\label{lomb}Results from Lomb-Scargle period analysis applied
to the EPIC/MOS (black) and EPIC/pn (blue) detrended light curves
while excluding the time interval between 8 and 22\,ks of elapsed time
to avoid contamination by the dip. A false alarm probability level of
0.3\% is marked by the horizontal line, and all peaks above this line
can be considered statistically significant at the 99.7\% (3-$\sigma$)
level. Two nearby significant frequency values $f_1$ and $f_2$
are marked in the top panel, together with the expected frequencies
of their first harmonics.
Separate periodograms calculated for the first 8\,ks of data
before the dip (part I) and those after 22\,ks (part II) have
been computed and are shown in the bottom two panels.
}
\end{figure}
 
 The results from the Lomb-Scargle procedure are shown in
Fig.~\ref{lomb}, derived from the detrended light curves from EPIC/MOS
(black) and EPIC/pn (blue). In the top panel, the initial result 
is shown with two significant frequencies of
$f_1 = 0.440\,\pm\,0.033$\,mHz and
$f_2 = 0.465\,\pm\,0.035$\,mHz. The expected location of the first
harmonic to each detected frequency is marked by a vertical dotted
line, and at most marginal evidence for harmonics is present.

 In addition, we have calculated two separate periodograms for 
two parts. Part I represents the first 8\,ks before the dip and
part II (22\,ks after the dip). The results are
shown in the bottom two panels of Fig.~\ref{lomb}. While the
first part is too short for any significant detection, the
second, longer, part clearly yields only a single frequency
of $0.448\,\pm\,0.036$\,mHz, corresponding to a period of
37.2 minutes.

\begin{figure}
\resizebox{\hsize}{!}{\includegraphics{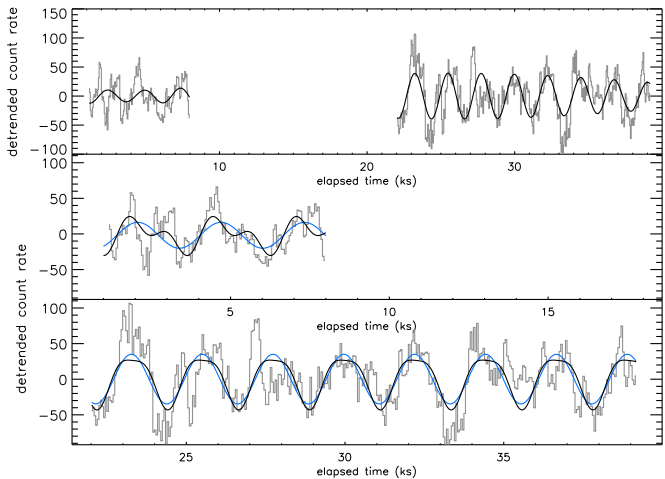}}
\caption{\label{sinfit}Top panel: Result from fitting two sine curves
to the detrended EPIC/pn light curve. The data are in light grey and
model in black. The strongest peak in the Lomb Scargle periodograms
are shown in the last two panels in Fig.~\ref{lomb} and as
sine curves overplotted over the first part
(middle) and second part (bottom). The blue curves are single
frequencies of 0.381\,mHz and 0.448\,mHz, and the black curves
are the sum of the ground frequency and the respective first
harmonic.
}
\end{figure}

 In order to investigate why the total light curve yields two
significant frequencies, we have performed several tests.
We applied the sine fitting method by \cite{dobrness09}, and
found two frequencies. However, a synthetic light curve modulated
with a single frequency, sampled and noised as the EPIC/pn data,
also yields two frequencies. We identify the reason to be
an increase in the amplitude of variations after the dip,
which mimics a beating cycle when fitted with two sine curves.
This is illustrated in Fig.~\ref{sinfit} where the detrended
EPIC/pn light curve is shown in grey with the best-fit 2-sine
curve overplotted. One can clearly see that the changes in
amplitude in the model reproduce the amplitude change in the
data. While the change in amplitude could be beating of
two signals, we consider it more likely that some other
process that is related to the dip is responsible for the
change in amplitude.

 For the discussion of the origin of the oscillations it
is important to know whether the first harmonic is also
present. In the bottom panels of Fig.~\ref{sinfit}, we show
the first and second parts with two different light curve
fits based on the frequencies detected in the respective
periodograms (see bottom two panels of Fig.~\ref{lomb}).
 The light blue
lines are single sine curves which were obtained after
fitting amplitudes and phase while keeping the frequencies
fixed at 0.381\,mHz for the first part and 0.448\,mHz for
the second part. The black curves are fits with fixed
fundamental frequencies plus the respective first harmonic
(twice the frequency values), also with variable
phase shifts and amplitudes. While the detection of
a first harmonic is in no case statistically significant
(as can be seen from Fig.~\ref{lomb}), the best fit
to the first part suggests that it may be present but is
not significant only for the reason that no more than three
cycles are covered.
We can therefore
not fully exclude that a first harmonic was present
before the dip which then disappeared
in the dip, as is clearly not present in the second
part.

 The frequency at $0.448\,\pm\,0.036$\,mHz (with a pulse
fraction of 7\% after the dip) seems clearly detected, but
we wish to caution that only seven cycles
are covered, bearing a certain risk of red noise. This
has been discussed in detail by \cite{vaughan10}
and references therein. We tested the confidence of
our signal using the method proposed by \cite{vaughan05}.
The 0.448-mHz peak is approximately at the 98\% confidence
limit. More data could mitigate this situation, but the
entire observation cannot be used
because the dip may have changed the conditions under
which a periodic signal is visible. The second
observation yields no periodic signal. Our conclusions
are based on seven apparently periodic
cycles in the second part of the first observation.

\section{Spectra}
\label{spectra}

\begin{figure*}
\resizebox{\hsize}{!}{\includegraphics{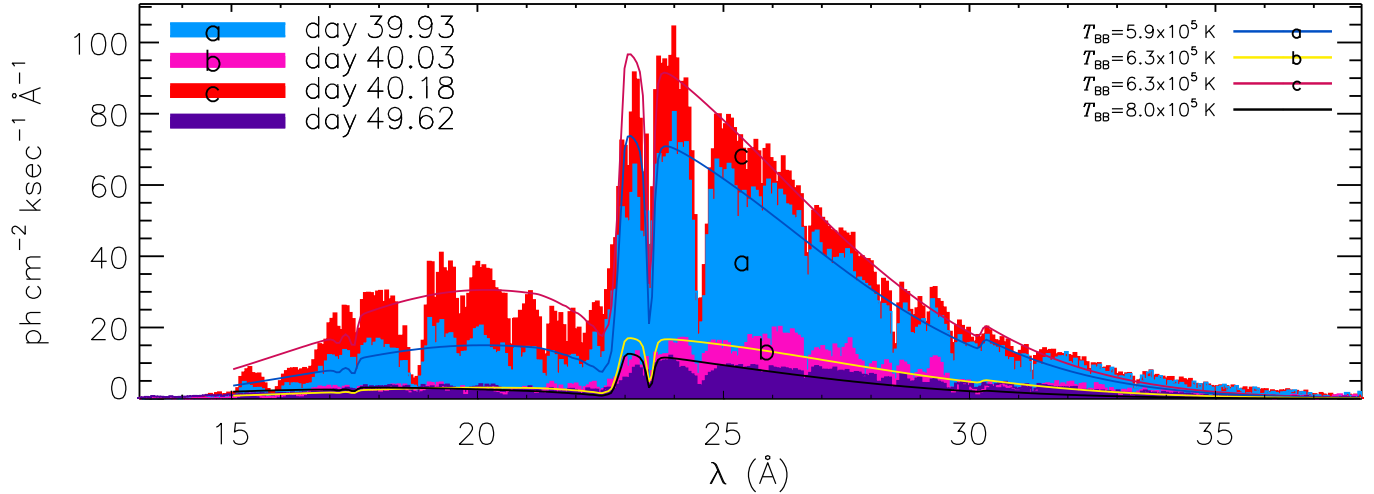}}
\caption{\label{spec}RGS spectra extracted from different epochs.
The time intervals over which each spectrum has been extracted
are marked in Fig.~\ref{lc} with the corresponding colors indicated
in the left legend (see Table~\ref{sepoch} for exposure details).
The units are flux units, and the spectra are
combined from RGS1 and RGS2 using the SAS tool {\tt rgsfluxer}.
The thin lines are best fit blackbody fits with the blackbody
temperatures given in the right legend.
}
\end{figure*}

The SAS tool {\tt rgsproc} was used to generate RGS events files
from the raw data and extract spectra in 0.01-\AA\ bins from a
standardized extraction region around the dispersed photons.
Time filters can be applied, allowing for the accumulation of spectra
over any given time interval. The SAS tool {\tt rgsfluxer} combines
all information from both spectrometers into a single spectrum
in photon flux units. It does not correct for redistribution
of monochromatic response into the dispersion channels.
Thus, the 'fluxed' spectrum is not suitable for quantitative
analyses but it can be used for the purpose of visualization
and comparison with broad-band models or other observations.

\begin{table*}
\begin{flushleft}
\caption{\label{sepoch}Separation of RGS spectra into four time segments with blackbody and atmosphere model parameters}
\vspace{-.4cm}
\begin{tabular}{lccr|rrrr|ccccccc}
\hline
Start & Color & $\Delta t^b$ & flux$^c$ & $T_{\rm BB}^d$ & $N_{\rm H}^{d,f}$& $R_{\rm BB}^d$ & $A_{\rm O}^{d,g}$& $T_{\rm eff}^e$ &$N_{\rm H}^{e,f}$ & $R_{\rm eff}^e$ & log($L_{\rm bol}$)$^e$ & $M_{\rm WD}^e$& $A_{\rm O}^{e,g}$\\
Date$^a$& Code & ks& &\multicolumn{3}{|r}{$10^5$\,K\hfill $10^8$\,cm}&&  $10^5$\,K &   & $10^8$\,cm    & erg\,s$^{-1}$ & M$_\odot$&\\
\hline
39.93 & blue (a) & 4.7  & 3.57 & 5.9 & 4.9 & 77&1.6 & $9.6  - 10.3$ & $2.1-2.4$ & $5.7-7.3$ & $38.42-38.52$ & $2.48-4.18$&$3.4-4.1$\\
40.03 & pink (b) & 6.1  & 0.98 & 6.3 & 4.1 & 22&2.0& $9.8  - 10.5$ & $1.6-1.8$ & $2.5-3.1$ & $37.73-37.81$ & $0.46-0.72$&$4.7-5.4$\\
40.18 & red (c)  & 16.5 & 5.09 & 6.3 & 4.7 & 57&1.2& $9.7  - 10.4$ & $2.0-2.3$& $5.8-7.8$ & $38.45-38.58$ & $2.54-4.34$&$2.4-2.8$\\
49.62 & purple   & 29.8 & 0.72 & 8.2 & 2.7 & 4&2.3& $10.3 - 10.5$ & $1.5-2.0$ & $1.9-2.2$ & $37.47-37.62$ & $0.27-0.37$&$4.4-4.7$\\
\hline
\end{tabular}

$^a$day after discovery, 2008, Apr. 10.7\\
$^b$Net Exposure time\\
$^c10^{-10}$\,erg\,cm$^{-2}$\,s$^{-1}$ over range 11-37\,\AA\\
$^d$from blackbody fit\\
$^e$from atmosphere model\\
$^f10^{21}$\,cm$^{-2}$\\
$^g$Oxygen abundance in {\tt TBabs} model, relative to solar by \cite{grev}
\end{flushleft}
\end{table*}

\subsection{Description}

In Fig.~\ref{spec}, the fluxed RGS spectra from four different
epochs are shown. The color code is the same as that used
in Fig.~\ref{lc}. In Table~\ref{sepoch}, the respective start
times in units of day after discovery, exposure time,
integrated X-ray band fluxes, and best-fit blackbody parameters
(see below) are listed. The fluxes are absorbed fluxes and
correspond to logarithmic luminosities of 30.67,
30.11, 30.82, and 29.98, respectively (in cgs units). The
spectra are all continuous spectra with deep absorption lines
at, e.g., $\sim 19$\,\AA\ and $\sim 24.5$\,\AA. The overall
shape of the spectra resembles that of an absorbed blackbody,
which is demonstrated by the overplotted best-fit blackbody
curves that were obtained by $\chi^2$ minimization. Spectral
bins around deep absorption lines have been discarded for better
reproduction of the continuum. The
effects of photoelectric absorption along the line of sight
were modeled using the Boulder-T\"ubingen model {\tt TBabs}
developed by \cite{wilms00} with variable oxygen abundance.
The best-fit parameters are listed in the last four columns
of Table~\ref{sepoch}, where the last column lists the oxygen
abundance in the {\tt TBabs} model, relative to solar
\citep{grev}. It was allowed to vary
in order to reproduce the depth of the O\,{\sc i} absorption edge
at $\sim 23$\,\AA. The corresponding 1s-2p absorption line
at 23.5\,\AA\ is also modeled as part of {\tt TBabs}.
The other absorption lines
are of photospheric origin, indicating that the spectrum
is that of an atmosphere. The deepest photospheric absorption
lines at $\sim 19$ and $\sim 24.5$\,\AA\ have been excluded
from the fit.

 With the given complex structure of the expanding nova
ejecta, the blackbody fits are only parameterizations.
As shall be shown in \S\ref{atmsect}, obtaining reliable results from
atmosphere modeling is sufficiently complex to require
a separate project, and the blackbody temperature is here
given only as a preliminary characterization of the spectral
hardness. The statistical uncertainties are small, of order
3\%, because of the high quality of the data, however the values
of reduced $\chi^2$ are 2.4, 1.8, 8.4, and 2.0 for days 39.93,
40.03, 40.18, and 49.62, respectively, and standard error
determination is not reliable for large values of reduced
$\chi^2$. Blackbody fits without discarding any spectral bins
yield systematically higher temperatures by $\sim 0.5\times10^5$\,K,
with much larger values of reduced $\chi^2$ of 6.3, 2.2, 28.8,
and 3.5. We emphasize that these temperatures
cannot be interpreted as the photospheric temperature because
a blackbody model does not account for the most fundamental physics in
the ejecta.

 For comparison with the observed UV fluxes listed in
Table~\ref{uv}, we have calculated the projected fluxes from the
blackbody fits.
For days 39.93, 40.03, 40.18, and 49.62, we found
(0.61, 1.5, and 2.1)$\times10^{-14}$\,erg\,cm$^{-2}$\,s$^{-1}$\,\AA$^{-1}$,
(0.06, 0.15, and 0.21)$\times10^{-14}$\,erg\,cm$^{-2}$\,s$^{-1}$\,\AA$^{-1}$,
(0.43, 1.1, and 1.5)$\times10^{-14}$\,erg\,cm$^{-2}$\,s$^{-1}$\,\AA$^{-1}$,
and
(1.9, 4.9, and 6.9)$\times10^{-17}$\,erg\,cm$^{-2}$\,s$^{-1}$\,\AA$^{-1}$
for the UVW1, UVM2, and UVW2 filters, respectively.
These fluxes are significantly lower than the measured values.
Since other sources of UV emission are likely present, these numbers do
not rule out a blackbody nature of the broad-band spectrum.

The changes in blackbody temperature are fairly small between
the three epochs of the first observation, which is consistent
with the findings by \cite{page09}. However, the second
observation yields a significantly harder spectrum. At the
same time, the best-fit value of $N_{\rm H}$ is of the same
order as the interstellar value derived from $E(B-V)$
measurements, indicating that the previous observations
contained excess absorption from local neutral material.
Note also that \cite{page09} found even lower
values of $N_{\rm H}$ around day 50 (see their fig.~4).
While the increases in spectral
hardness could be due to an increasing effective
temperature, which would be consistent with a smaller
effective blackbody radius, changes in the atmospheric
structure may also have occurred. Also, an increasing
degree of ionization of circumbinary neutral material can
lead to spectral hardening as demonstrated by \cite{ness_rsoph}
for RS\,Oph. The degree of ionization of oxygen influences the
depth of the O\,{\sc i} absorption edge at 23\,\AA\ and the
associated 1s-2p line at 23.5\,\AA. The parameter $A$(O), the
abundance of neutral oxygen in the cold absorber component,
yields higher values for a lower degree of ionization
because more neutral oxygen is present. The results given
in the last column of Table~\ref{sepoch} indicate that on day
40.03, when the total flux was lower, the degree of ionization
was also lower, which can be explained by recombination in a
photoionized plasma that was less irradiated during the times
of low-flux emission.

 Some model-independent conclusions can be drawn
by direct inspection of the details in the spectra.
Trends of abundances can be estimated from line depths,
and the local expansion velocity can be determined from
line shifts on a quantitative level.

\subsection{Details in the High-resolution spectra}

\begin{figure*}
\resizebox{\hsize}{!}{\includegraphics{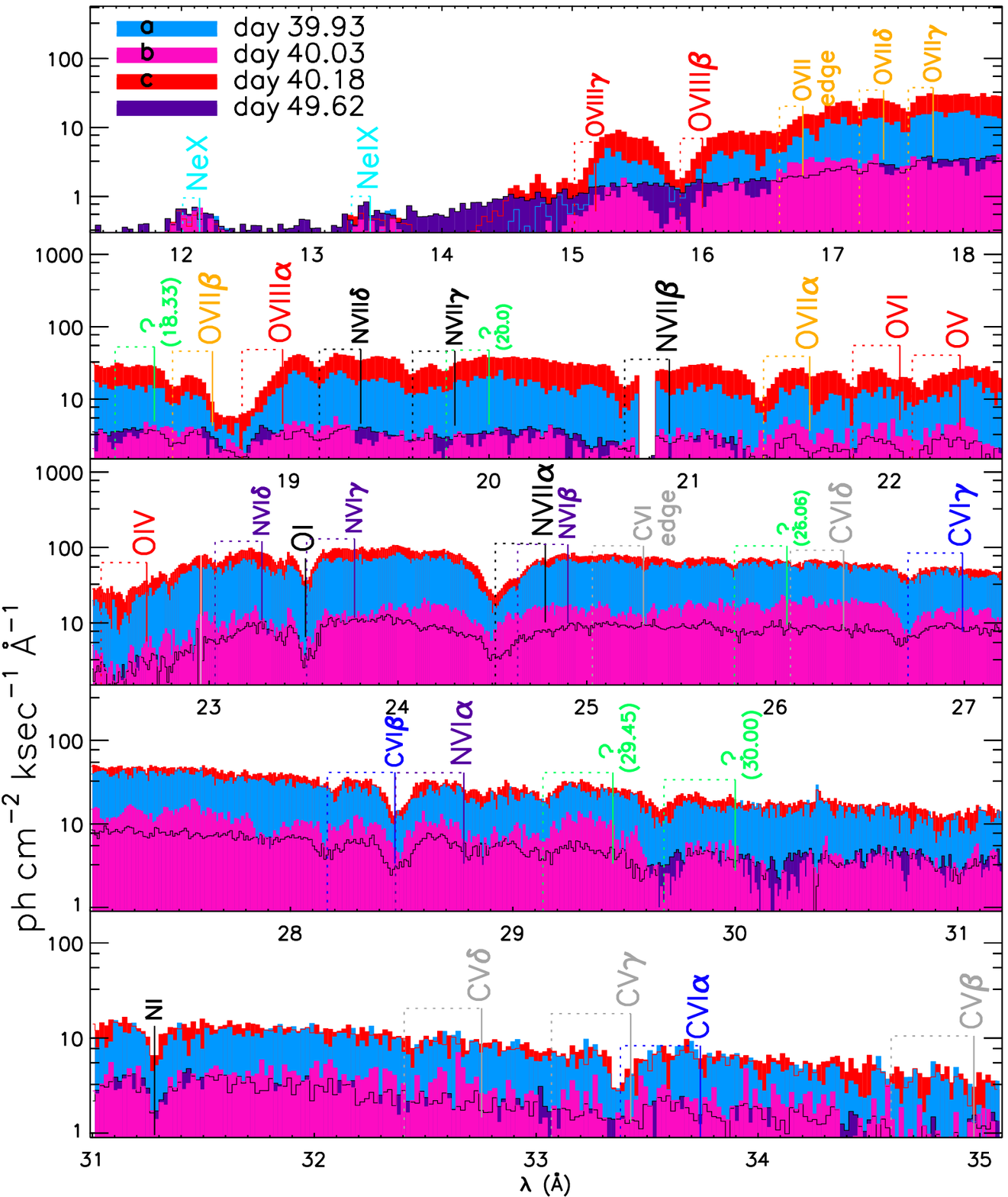}}
\caption{\label{levol}Combined 'fluxed' RGS1 and RGS2 spectra of
V2491\,Cyg extracted from different time intervals. The start
times are given in the upper left legend, and the colors correspond
to those used in Fig.~\ref{lc} (see Table~\ref{sepoch}). Important
absorption lines are marked with small
vertical solid lines at their rest wavelengths and with dashed lines
at the wavelengths corresponding to blue-shifts of
$v=-3200$\,km\,s$^{-1}$. The line labels are color coded indicating
lines of N\,{\sc vii} in black,  N\,{\sc vi} in purple, O\,{\sc iv-vi}
and O\,{\sc viii} in red, O\,{\sc vii} in orange, carbon lines in blue,
interstellar lines in black, and unidentified lines in green.
Labels in grey indicate lines that are not detected.
}
\end{figure*}

 The details of the spectra are shown in Fig.~\ref{levol},
using the same color code as in Figs.~\ref{lc} and \ref{spec}.
Line labels of prominent transitions are placed at their rest
wavelengths using different colors that
indicate groups of transitions and elements. The groups are:
neutral (interstellar) lines (black), the H-like Ly series (1s-$n$p)
of N\,{\sc vii}, O\,{\sc viii}, and C\,{\sc vi} (black, red, and
blue, respectively), the He-like 1s-$n$p series of N\,{\sc vi}, O\,{\sc vii},
and C\,{\sc v} (purple, orange, and grey), Neon emission lines
(cyan), and absorption features that cannot be identified
(green). Lines that are not detected but are of interest
are marked in light grey (e.g., C\,{\sc v}).
 Expected blue shifts corresponding to a Doppler velocity of
$-3200$\,km\,s$^{-1}$ are indicated by dotted lines for each
transition. For the
majority of lines, the dotted lines coincide with deep absorption
troughs. Except for the two interstellar lines of O\,{\sc i}
(23.5\,\AA) and N\,{\sc i} (31.28\,\AA), all lines indicate
roughly the same expansion velocity of $-3200$\,km\,s$^{-1}$.

In the spectrum representing the low emission phase during the
first observation (phase b, pink), the N\,{\sc vii} and the
C\,{\sc vi} Ly$\alpha$ lines at $\lambda_0=24.78$\,\AA\ and
$\lambda_0=33.74$\,\AA, respectively, appear to be
blue shifted by a larger amount than in the other two
spectra from the first observation (phases a and c in blue and
red). Meanwhile, the Ly$\beta$ lines of the same elements and the
N\,{\sc vi} lines show the same blue shift.

\begin{figure}
\resizebox{\hsize}{!}{\includegraphics{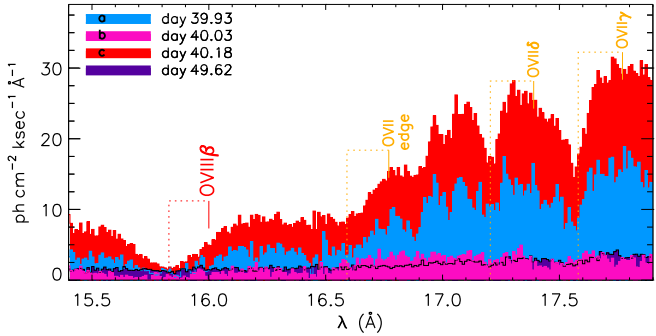}}
\caption{\label{o7}Same as Fig.~\ref{levol} in linear
units, focusing on the O\,{\sc vii} edge at 16.77\,\AA.
}
\end{figure}

 The H-like Ly series lines of N\,{\sc vii} can be identified
up to 1s-5p (Ly$\delta$). The recombination/ionization continuum
of N\,{\sc vii} is expected at 18.59\,\AA\ (667.1\,eV) but is
not seen (and is not marked) because it overlaps with
the O\,{\sc vii} 1s-3p line at 18.67\,\AA.
 For O\,{\sc viii}, the 1s-$n$p series can be seen
up to $n=4$ (Ly$\gamma$). At shorter wavelengths, not enough
continuum is present to detect higher-$n$ Ly series lines of
O\,{\sc viii}.
For C\,{\sc vi}, the 1s-$n$p series can only be seen up to
Ly$\gamma$, while Ly$\delta$ and the recombination/ionization
continuum at 25.3\,\AA\ (490\,eV) are not present.
The He-like series lines of N\,{\sc vi} and O\,{\sc vii} are
detected up to 1s-5p, and for O\,{\sc vii}, the
recombination/ionization continuum at 16.77\,\AA\ (739.3\,eV)
is present between 16.6-16.8\,\AA. This is shown in more detail
in Fig.~\ref{o7}. The depth of the line is strongest in
the spectrum from day 40.18. The shape of the edge indicates an
imbalance in favor of ionization, thus O\,{\sc viii} is
being pumped by photoionization. The N\,{\sc vi} recombination/ionization
continuum at 22.46\,\AA\ (552.1\,eV) cannot be detected
because it overlaps with a number of low-ionization oxygen lines,
O\,{\sc vi}, O\,{\sc v}, and O\,{\sc iv}, marked in red.
While the He-like 1s-2p line of C\,{\sc v} (40.27\,\AA) is outside
the spectral range of the RGS, the high-order series lines are
in the RGS band pass and are not detected (grey labels).

Unidentified absorption features
are marked with a question mark at a wavelength that
would be the rest wavelength if that feature was photospheric and
thus blue shifted by the same amount as the other lines, but they
could also be interstellar lines. Identification of these lines
might be possible if they were also seen in other nova spectra with
a different velocity profile (see \S\ref{cmpothernovae}). Seeing similar
features at the same wavelengths
would be an argument in favor of interstellar lines, while line
shifts by an amount commensurate with other line shifts would
be more suggestive of photospheric lines.

 At 12 and 13.5\,\AA, the H-like and He-like lines of
Ne\,{\sc x} and Ne\,{\sc ix} are clearly detected in emission.
On day 49.62, this wavelength region is dominated by continuum
emission originating from an increased Wien tail of the photospheric
continuum spectrum. In the logarithmic plot it is difficult to see,
but the Ne lines are still significantly present, albeit with a
lower flux \citep{v2491_xmm2}.
 
\subsection{Line profile modeling}
\label{lprofil}

 The deepest absorption line is the Ly$\alpha$ line of
O\,{\sc viii} at 18.8\,\AA\ ($\lambda_0=$18.97\,\AA).
\cite{ness09}
found that it was saturated without going to zero. This is the
only line for which the bottom of the line profile is flat,
indicating that it has the largest column density and thus
the highest number density. In addition to a high
oxygen abundance, the temperature structure at the
time of the observation provides particularly advantageous
conditions for the formation of the O\,{\sc viii} ionization
state. The latter is supported by the presence of the
O\,{\sc vii} to O\,{\sc viii} ionization edge at 16.8\,\AA\
(see Fig.~\ref{o7}), which is an indicator for photoionization.
 
 We have used a slightly modified version of the method
described by \cite{ness09} to determine line shifts, widths,
and optical depths at line center for each of the four spectra
from days 39.93, 40.03, 40.18, and 49.62. A line profile model
spectrum is calculated over a narrow wavelength range around an
absorption line, based on equation 1 in \cite{ness09}.
For faster computation, we perform no correction for
interstellar absorption (thus $T(\lambda)\equiv 1$),
and for the continuum we assume a linear function rather
than a blackbody curve. These two effects are marginal
over the small wavelength region around the absorption
lines. We have compared the results by \cite{ness09}
with the same spectra and lines but with the modified
method and get consistent results.

 The resulting parameters of interest are the optical depth
at line center, $\tau_{\rm c}$, the central (rest) wavelength,
$\lambda_0$, and the Gaussian line width, $\sigma$.
The line shifts, $(\lambda-\lambda_0)$, and line
widths were converted to Doppler velocities
$v_{\rm shift}={\rm c}(\lambda-\lambda_0)/\lambda_0$
and $v_{\rm width}={\rm c}\sigma/\lambda_0$,
where c is the speed of light.
The instrumental line broadening can be estimated as
$v_{\rm instr}\sim 300$\,km\,s$^{-1}$ and was accounted for in
computing $v_{\rm width}=\sqrt{ ({\rm c}\sigma/\lambda_0)^2
-v_{\rm instr}^2}$.

In addition to these parameters, line column
densities, $N_{\rm X}$, were obtained by integration over the
opacity $\tau(\lambda)$ as defined in equation 2 of \cite{ness09},
using oscillator strengths extracted from
the Chianti database \citep{landi06}, which is an
approximation for strong lines.

\begin{figure}
\resizebox{\hsize}{!}{\includegraphics{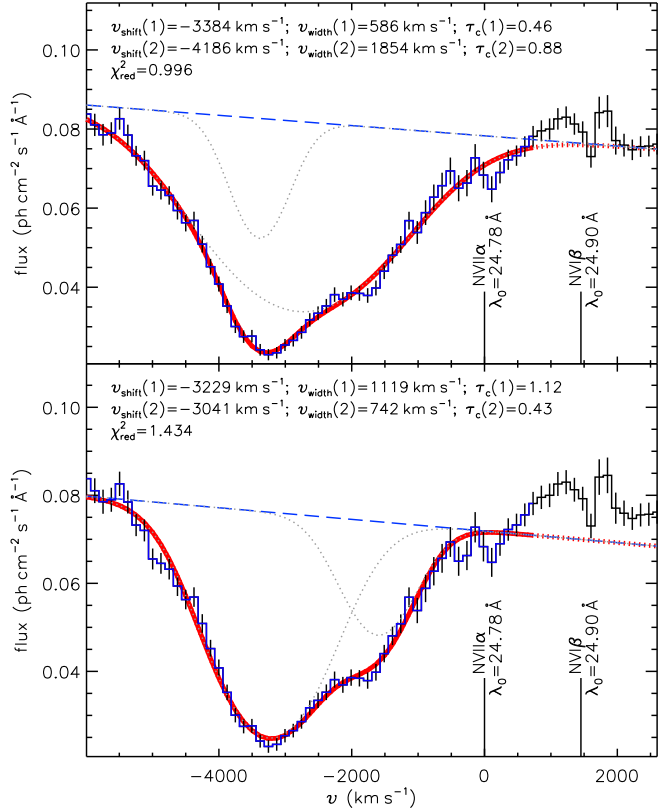}}
\caption{\label{n7}Line profile fitting to the spectrum of day 40.18,
following a modified method by (Ness 2010, see text) for the two
blended lines of N\,{\sc vii} ($\lambda_0=24.78$\,\AA) and
N\,{\sc vi} ($\lambda_0=24.90$\,\AA). The observed spectrum is plotted
in histogram style (blue for fitted part and black for non-fitted part),
the best-fit model with a thick red line (solid and dotted for fitted
and non-fitted, respectively), and the model components for each line
with grey dotted lines. While the fit in the top panel yields lower
$\chi^2$, we consider the fit in the bottom panel as the one
with better physical motivation because the parameters are closer to
the values from the
other spectra (see Table~\ref{linetab}), and a dip between 
$-2000$ and $-1800$\,km\,s$^{-1}$ is better reproduced.
}
\end{figure}

 All results are summarized in Table~\ref{linetab}, where the top
block lists the results for the two brightest spectra
taken before and after the dip in the first observation (blue and red
shadings in the top panel of Fig.~\ref{lc}), and the bottom block
represents the two faintest spectra represented by pink and
purple colors in Fig.~\ref{levol}. Given are name of ion, ionization
parameter (see last paragraph of this subsection), rest wavelength,
oscillator strength, line
shift and width, optical depth at line center, and line column
density. The given errors were obtained for each parameter from a
grid of parameter values versus corresponding $\chi^2$ by
interpolating the two parameter values on each side of the minimum
that are larger by 3.53 than the minimum $\chi^2$. This increase in
$\chi^2$, with three free parameters of interest, yields 1-$\sigma$
uncertainties.
While stepping through the grid, all other parameters were iterated,
including the parameters for the continuum that are statistically
'uninteresting' \cite[see][]{avni76},
starting at their best-fit values. If a better fit was found during
this procedure, a new minimization was started with the new parameters
until the fit was stable. With this procedure, we are confident to
have found global minima in $\chi^2$. Only in one case have we not
accepted the best fit, which is illustrated in Fig.~\ref{n7}, where
in the top panel the global best fit is given while in the bottom panel
our preferred fit is shown, which is a local minimum in $\chi^2$. The
best fit requires a broad N\,{\sc vi} line and a narrow N\,{\sc vii} line,
while all other observations yield the opposite result. This single
result does not agree with the other results. We illustrate this
to make the point that rigorous fitting does not always lead to the
best results. In Fig.~\ref{n7}, a dip between $-2000$ and
$-1800$\,km\,s$^{-1}$ can be seen which could well be originating
from the N\,{\sc vi} transition. In the bottom panel, this dip is much better
reproduced, yielding a velocity shift for N\,{\sc vi} that is
consistent with that of the large majority of other absorption
lines. Although this fit has a higher value of $\chi^2$, we prefer
this fit over the one in the top panel, and in Table~\ref{linetab},
the best-fit values are given.

\begin{table*}
\begin{flushleft}
\caption{\label{linetab}Results from line profiles measurements}
\begin{tabular}{p{.6cm}p{.4cm}p{.5cm}p{.3cm}cccc|cp{1.2cm}p{1.5cm}c}
\hline
\multicolumn{2}{l}{Ion \hfill log($\xi$)} & $\lambda_0$ & $f$ & $v_{\rm shift}$ & $v_{\rm width}$ & $\tau_{\rm c}$ & $N_{\rm X}$ & $v_{\rm shift}$ & $v_{\rm width}$ & $\tau_{\rm c}$ & $N_{\rm X}$\\
\multicolumn{2}{r}{(erg\,s$^{-1}$\,cm)}& \AA & & km\,s$^{-1}$ & km\,s$^{-1}$ & & $10^{16}$\,cm$^{-2}$ & km\,s$^{-1}$ & km\,s$^{-1}$ & & $10^{16}$\,cm$^{-2}$ \\
\hline
\multicolumn{8}{l|}{\bf Day\,39.93}&\multicolumn{4}{l}{\bf Day\,40.18}\\
C\,{\sc vi}$\beta$ & 1.30 & 28.46 & 0.16 & $-2705^{+146}_{-172}$ & $<353$ & $0.37^{+51.01}_{-0.35}$ & $<1.41$ & \mbox{$-3034^{+131}_{-135}$} & \mbox{$687^{+280}_{-213}$} & \mbox{$0.29\pm 0.09$} & \mbox{$0.27\pm 0.09$}\\
C\,{\sc vi}$\alpha$ & 1.30 & 33.74 & 0.83 & $-3010^{+230}_{-220}$ & $626^{+313}_{-197}$ & $0.94^{+0.57}_{-0.37}$ & $0.12^{+0.08}_{-0.05}$ & \mbox{$-3213^{+113}_{-90}$} & \mbox{$<549$} & \mbox{$0.90^{+0.35}_{-0.24}$} & \mbox{$0.12^{+0.05}_{-0.03}$}\\
N\,{\sc vi}$\beta$ & 0.80 & 24.90 & 0.14 & $-3157^{+140}_{-213}$ & $567^{+292}_{-191}$ & $0.46\pm 0.15$ & $0.62\pm 0.21$ & \mbox{($-3041\pm 57$)$^\star$} & \mbox{($726^{+79}_{-72}$)$^\star$} & \mbox{($0.43\pm 0.03$)$^\star$} & \mbox{($0.57\pm 0.05$)$^\star$}\\
N\,{\sc vi}$\alpha$ & 0.80 & 28.78 & 0.66 & $-3119^{+94}_{-87}$ & $631^{+150}_{-116}$ & $0.87\pm 0.20$ & $0.19\pm 0.05$ & \mbox{$-3147\pm 40$} & \mbox{$611^{+58}_{-54}$} & \mbox{$0.90\pm 0.09$} & \mbox{$0.20\pm 0.02$}\\
N\,{\sc vii}$\alpha$ & 1.60 & 24.78 & 0.83 & $-3275^{+107}_{-143}$ & $1146^{+144}_{-139}$ & $1.30\pm 0.10$ & $0.29\pm 0.03$ & \mbox{($-3229\pm 27$)$^\star$} & \mbox{($1108^{+29}_{-28}$)$^\star$} & \mbox{($1.12\pm 0.03$)$^\star$} & \mbox{($0.25\pm 0.01$)$^\star$}\\
O\,{\sc vii}$\beta$ & 1.20 & 18.63 & 0.15 & $-3227^{+150}_{-136}$ & $777^{+263}_{-187}$ & $0.78\pm 0.31$ & $1.84\pm 0.82$ & \mbox{$-3178\pm 58$} & \mbox{$965^{+95}_{-90}$} & \mbox{$0.91\pm 0.08$} & \mbox{$2.08\pm 0.23$}\\
O\,{\sc vii}$\alpha$ & 1.20 & 21.60 & 0.68 & $-3175^{+186}_{-193}$ & $716^{+290}_{-200}$ & $0.81^{+0.34}_{-0.25}$ & $0.31^{+0.14}_{-0.11}$ & \mbox{$-3308\pm 64$} & \mbox{$855^{+100}_{-87}$} & \mbox{$0.95\pm 0.11$} & \mbox{$0.35\pm 0.05$}\\
O\,{\sc viii}$\beta$ & 2.70 & 16.00 & 0.16 & $-3292^{+319}_{-324}$ & $2752^{+613}_{-493}$ & $1.30\pm 0.29$ & $3.47\pm 1.08$ & \mbox{$-3275\pm 107$} & \mbox{$2825^{+205}_{-188}$} & \mbox{$1.41\pm 0.09$} & \mbox{$3.52\pm 0.36$}\\
O\,{\sc viii}$\alpha$ & 2.70 & 18.97 & 0.83 & $-3703\pm 107$ & $2545^{+361}_{-283}$ & $2.37\pm 1.25$ & $0.84^{+0.53}_{-0.41}$ & \mbox{$-3732\pm 41$} & \mbox{$2302^{+111}_{-97}$} & \mbox{$3.24\pm 0.52$} & \mbox{$1.13\pm 0.22$}\\
\hline
\multicolumn{8}{l|}{\bf Day\,40.03}&\multicolumn{4}{l}{\bf Day\,49.62}\\
C\,{\sc vi}$\beta$ & 1.30 & 28.46 & 0.16 & $-3235^{+237}_{-174}$ & $<487$ & $0.54^{+5.29}_{-0.46}$ & $<7.23$ & $-2996^{+201}_{-192}$ & $558^{+226}_{-178}$ & $0.36^{+0.18}_{-0.14}$ & $0.33^{+0.18}_{-0.14}$\\
C\,{\sc vi}$\alpha$ & 1.30 & 33.74 & 0.83 & $-3603^{+236}_{-155}$ & $<561$ & $1.57^{+34.13}_{-0.59}$ & $<1.03$ & -- & -- & -- & --\\
N\,{\sc vi}$\beta$ & 0.80 & 24.90 & 0.14 & $-3044^{+293}_{-316}$ & $398^{+553}_{-398}$ & $0.23^{+0.69}_{-0.15}$ & $0.31^{+1.17}_{-0.21}$ & $-2516^{+205}_{-210}$ & $<920$ & $0.19^{+0.16}_{-0.13}$ & $0.25^{+0.25}_{-0.18}$\\
N\,{\sc vi}$\alpha$ & 0.80 & 28.78 & 0.66 & $-3106\pm 129$ & $364^{+194}_{-364}$ & $0.72^{+0.39}_{-0.26}$ & $0.16^{+0.10}_{-0.06}$ & $-3165\pm 76$ & $656^{+101}_{-85}$ & $0.82\pm 0.15$ & $0.18\pm 0.04$\\
N\,{\sc vii}$\alpha$ & 1.60 & 24.78 & 0.83 & $-3661^{+170}_{-144}$ & $1230^{+349}_{-254}$ & $0.76\pm 0.14$ & $0.17\pm 0.04$ & $-3029\pm 85$ & $1150^{+176}_{-143}$ & $0.79\pm 0.09$ & $0.17\pm 0.03$\\
O\,{\sc vii}$\beta$ & 1.20 & 18.63 & 0.15 & $-2844^{+357}_{-583}$ & $<7903$ & $0.59^{+10.63}_{-0.40}$ & $<22.10$ & $-3053\pm 144$ & $766^{+206}_{-156}$ & $1.19^{+0.50}_{-0.34}$ & $2.84^{+1.38}_{-0.91}$\\
O\,{\sc vii}$\alpha$ & 1.20 & 21.60 & 0.68 & -- & -- & -- & -- & $-3327\pm 174$ & $777^{+249}_{-202}$ & $0.79\pm 0.29$ & $<0.43$\\
O\,{\sc viii}$\beta$ & 2.70 & 16.00 & 0.16 & $-3964^{+536}_{-544}$ & $4263^{+1296}_{-878}$ & $1.43\pm 0.36$ & $3.81\pm 1.34$ & $-3137^{+422}_{-477}$ & $894^{+711}_{-586}$ & $0.28^{+0.33}_{-0.15}$ & $0.82^{+1.08}_{-0.49}$\\
O\,{\sc viii}$\alpha$ & 2.70 & 18.97 & 0.83 & $-3721^{+316}_{-337}$ & $2360^{+608}_{-461}$ & $2.16^{+1.20}_{-0.63}$ & $0.81^{+0.55}_{-0.29}$ & $-3776\pm 102$ & $908^{+163}_{-167}$ & $5.85^{+9.17}_{-2.32}$ & $<13.06$\\
\hline
\end{tabular}
$^\star$Uncorrelated errors given, see text for details
\end{flushleft}
\end{table*}

\begin{figure}
\resizebox{\hsize}{!}{\includegraphics{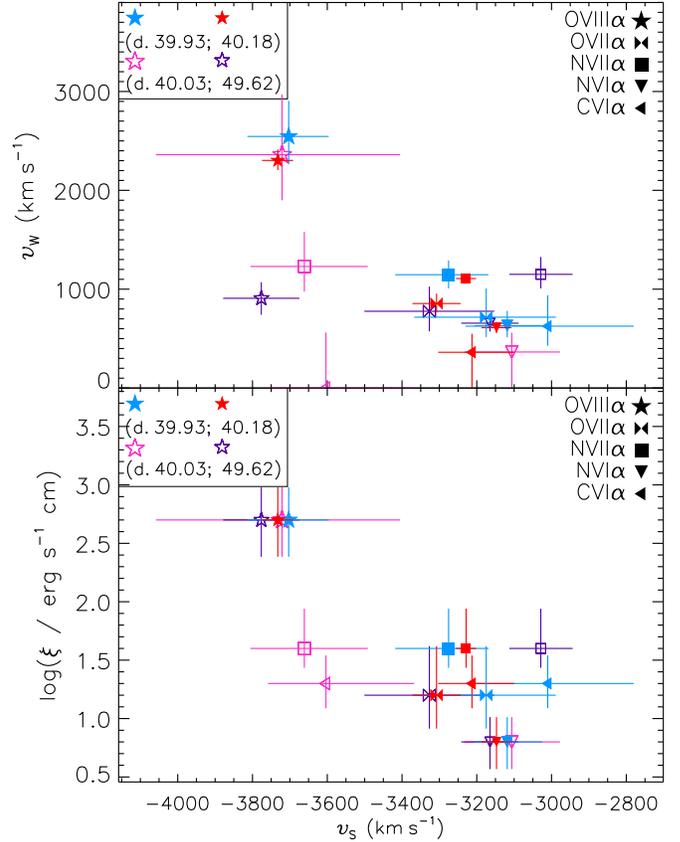}}
\caption{\label{shifts}Line shifts plotted versus line widths (top) and
versus ionization parameter yielding the highest column density (bottom),
with the values
given in Table~\ref{linetab}. The symbols indicate different lines as
indicated in the right legend and four symbol colors correspond to the
four epochs of extracted spectra (Fig.~\ref{lc} and Table~\ref{sepoch}).
}
\end{figure} 

 Fig.~\ref{shifts} is a graphical illustration of the line shifts and
line widths (top panel) for different lines at different times. Only
strong 1s-2p (Ly$\alpha$ and He$\alpha$) transitions were selected
from Table~\ref{linetab} for this plot for better clarity.
The majority of lines are blue shifted by $-3000$ to $-3400$\,km\,s$^{-1}$
and are broadened by up to 1200\,km\,s$^{-1}$, which is significantly
broader than the expected line width from instrumental broadening. We
can thus conclude that the ejecta are significantly extended, allowing
us to view a range of expansion velocities and thus a range of
different plasma layers.

The O\,{\sc viii} line
clearly stands out with consistently larger line shifts and widths in
all three data sets of the first observation. By day 49.6, the line has
become narrower but has the same line shift.

The N\,{\sc vii} and C\,{\sc vi} lines are blue shifted
by a larger amount during the dip in the first observation (day
40.03 in pink). Together with the O\,{\sc viii} line, these are
all H-like transitions while the He-like N\,{\sc vi} line, which
is formed at lower temperatures, is not more blue shifted during the
dip.
This could indicate that the high-temperature lines probe higher wind
velocities which are likely found further outside.

 To get a better idea about systematic temperature effects, we
computed a characteristic ionization parameter $\xi=L/(nr^2)$,
with $L$ the bolometric luminosity of the ionizing source in the
$1-1000$ Rydberg range, $n$ the electron density of the
photo-ionized absorbing gas, and $r$ the distance from the
ionizing source. For a grid of ionization parameters, column
densities have been computed for each ion using ionization-balance
calculations done with Cloudy \citep{cloudy}. The spectral energy
distribution (SED) input into Cloudy consists of IR data from
\cite{naik09}, optical data from the AAVSO database presented by
\cite{hachisu09}, UV and soft
X-ray fluxes from this paper, and hard X-ray fluxes from
\cite{page09}. Solar abundances by \cite{lodders09} were
assumed. A description of how this was done can be
found in \S 6.2 of the SPEX
manual\footnote{www.sron.nl/files/HEA/SPEX/manuals/manual.pdf}
\citep{spex}.
From the resulting balance curves, the ionization parameter yielding
the largest absorption column density for each ion is selected for
plotting along the vertical axis in the bottom panel of
Fig.~\ref{shifts}. As a confidence range, the range of ionization
parameter is chosen that includes all column densities greater than
10\% below the value at the peak. A trend seems suggestive that
lines with higher ionization parameter are more blue shifted, but
this conclusion depends very much on the measurements of the
O\,{\sc viii} line. With the given set of lines, it is thus difficult
to establish a trend with ionization temperature.
A more systematic approach will be presented in Pinto et al. (in
preparation).

\subsection{Comparison with atmosphere models}
\label{atmsect}

\begin{figure}
\resizebox{\hsize}{!}{\includegraphics{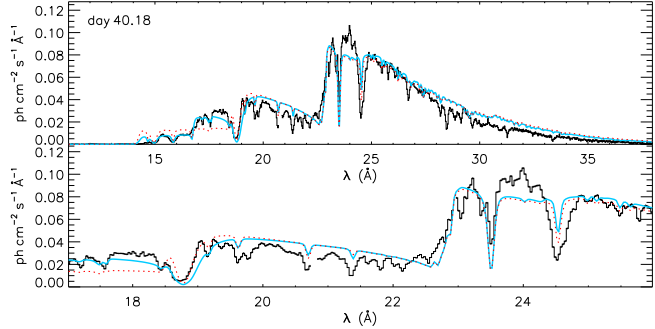}}
\caption{\label{atm}Comparison between model atmosphere and RGS spectrum
of V2491\,Cyg commencing on day 40.18. Models are taken from
\cite{atmtables}. Their table~1 lists different abundance categories,
and the blue solid and red dotted lines are models \#007 and \#003,
respectively. Both models are blue shifted by $z=-0.01$,
corresponding to $v_{\rm shift}=-3000$\,km\,s$^{-1}$.
All models assume log\,$g=9$.
}
\end{figure}

 The closest physical approach to derive global parameters of the nova ejecta
are non-LTE atmosphere models. Of particular interest for understanding nova
evolution are the effective temperature and the mass of the underlying WD.
Here, we use the publicly available tabulated non-LTE atmosphere models
described by \cite{atmtables}. The T\"ubingen Model Atmosphere Package
(TMAP) has been developed to study atmospheres of hot white dwarfs,
where assumptions such as hydrostatic equilibrium and a plane parallel
geometry are valid approximations.
In \S\ref{atmdisc} we discuss in more depth the limitation of
these assumptions when applied to novae.

 We used all available data tables and obtained best fits with
{\tt xspec} \citep{xspec} for all four spectra extracted from the
time intervals listed in Table~\ref{sepoch}. We carried out simultaneous
fits to the separate RGS1 and RGS2 spectra and used the fluxed spectra
only for plotting. We corrected for photoelectric absorption using the
{\tt TBnew} module developed by \cite[][Wilms, Juett, Schulz, Nowak,
in preparation]{wilms00,wilms06} with a variable column density of
neutral hydrogen, $N_{\rm H}$, and abundance of neutral oxygen.

 The public database currently contains only models assuming log\,$g=9$,
covering a range of chemical abundances. In Fig.~\ref{atm},
the best fit models \#007 and \#003 are shown with blue solid and red dotted
lines, respectively, in comparison to the spectrum from day 40.18.
The top panel shows the entire spectrum, and the overall reproduction
of the data appears reasonable. The oxygen-rich
Model \#007 performs better at shorter wavelengths, supporting our
earlier conclusion of a high O abundance. In the bottom panel, a
zoom-in is shown, illustrating that some of the absorption lines are
represented well.
Values of reduced $\chi^2$ (number of degrees of freedom in brackets)
for model \#007 are 3.7 (5511), 2.2 (5613), 21.8 (5616), and 
4.5 (5504), for days 39.93, 40.03, 40.18, and 49.62, respectively.
Note that the number of degrees of freedom is overestimated (and thus
reduced $\chi^2$ underestimated) because a large number of parameters
in the models are not counted as free parameters because they are not
varied. Hard-coded values such as NLTE parameters have already been
chosen in an early stage of model development and are not further
considered during routine fitting, but they would likely change the
results if modified.
Therefore, they are not necessarily 'uninteresting' parameters
in the sense of \cite{avni76} and may be of importance in deeper
studies involving new computations for the special situation of
an individual system such as V2491\,Cyg. In particular, different
values of log\,$g=9$ that are currently not publicly available have
to be considered, which will significantly increase the parameter
space, likely yielding different results.

 The model parameters are listed in the right part of
Table~\ref{sepoch}, where the ranges include the best-fit results
from using all available models. The ranges do not reflect the
full accuracy of physical quantaties but only the ranges of parameters
if different abundance classes are assumed. It is impossible to predict
how the parameters would change if effects of expansion or extended
geometry were included. Radii were computed from the model
normalization, $R=10^{-11}\times \sqrt{\rm norm}\times d$, with
$d=10.5$\,kpc the distance, bolometric luminosities from the
Stefan-Boltzmann law, and masses from the radii and log\,$g=9$,
$M=R^2 g/$G with the gravitational constant
G$=6.674281\times10^{-8}$ in cgs units.

The parameter of highest interest is the effective temperature, for
which we find the same trend with time
as already found from the blackbody fits. With the result from the line
profiles that the ejecta are extended, allowing us to view through
different velocity layers, it is plausible that we are also viewing
different temperature layers in the same spectra. Therefore, the
definition of a single characteristic temperature does not have
the same physical meaning as in a white dwarf evolutionary model
and can not be used in the same way. We also caution that the underlying
parameter log\,$g=9$ is unrealistic for nova ejecta. The derived masses
are completely unrealistic with some values much larger than the
Chandrasekhar mass limit. We note that standard mass-radius relations
do not apply for active novae since the radiative energy is not
drawn from gravitational contraction but from nuclear burning.
Also, the bolometric luminosity computed from the
Stefan-Boltzmann law, assuming the same radius, is of order
$10^{38}$\,erg\,s$^{-1}$ and thus also unrealistically high for the
late stage of evolution.

  It has to be kept in mind that
the underlying model assumption is a plane parallel geometry, and
strictly speaking, the calculation of a radius is in violation to this
model assumption. Therefore, all calculations of WD masses and
luminosities that have been obtained from hydrostatic plane parallel
models have to be treated with greatest caution.

 With all these concerns in mind, these models represent the data
surprisingly well as can be seen in Fig.~\ref{atm} making
it tempting to trust this approach.
Possibly, new data tables with lower log\,$g$ may yield more realistic
masses, however, spectral fits of models with more realistic
parameter combinations have not so far yielded better fits than
can be seen in Fig.~\ref{atm} \cite[see, e.g., ][]{nelson07,rauch10}.

We argue that a much more detailed discussion of refinement
of atmosphere models is required, which is beyond the scope of this paper. 
While the data tables can be extended to include a larger range of log\,$g$,
it will in our view be unavoidable that individual models are calculated
for each nova, accounting for the individual conditions. Furthermore, more
studies are needed to investigate the effects on the derived effective
temperature when accounting for the extended geometry and the expansion.

\begin{table}
\begin{flushleft}
\caption{\label{unidentified}List of unidentified lines}
\begin{tabular}{lllll}
\hline
$\lambda_{\rm proj}^a$ (\AA) & \multicolumn{3}{c}{$\lambda_{\rm obs}$ (\AA)}& comment\\
&V2491\,Cyg & V4743\,Sgr & RS\,Oph&\\
\hline
20.00 & 19.80 & -- & --  & low-$T$ line\\
25.70 & -- & -- &25.6\\
25.80 & -- & -- & 25.7\\
26.06 & 25.78 & 25.85 & 25.90 & low-$T$ line\\
26.21 & -- & -- & 26.1\\
26.93 & -- & -- & 26.83\\
27.41 & -- & -- & 27.3\\
27.61 & -- & -- & 27.5\\
27.71 & -- & -- & 27.6\\
28.11 & -- & -- & 28.0\\
29.45 & 29.13 & 29.18 & 29.30 & high-$T$ line\\
30.00 & 29.66 & -- & 29.70 & high-$T$ line\\
30.42 & -- & -- & 30.3\\
\hline

\end{tabular}

$^a$Projected, =assumed rest wavelength, assuming blue shifts of
3300\,km\,s$^{-1}$ for V2491\,Cyg, 2400\,km\,s$^{-1}$ for V4743\,Sgr,
and 1200\,km\,s$^{-1}$ for RS\,Oph.
\renewcommand{\arraystretch}{1}
\end{flushleft}
\end{table}

\begin{figure}
\resizebox{\hsize}{!}{\includegraphics{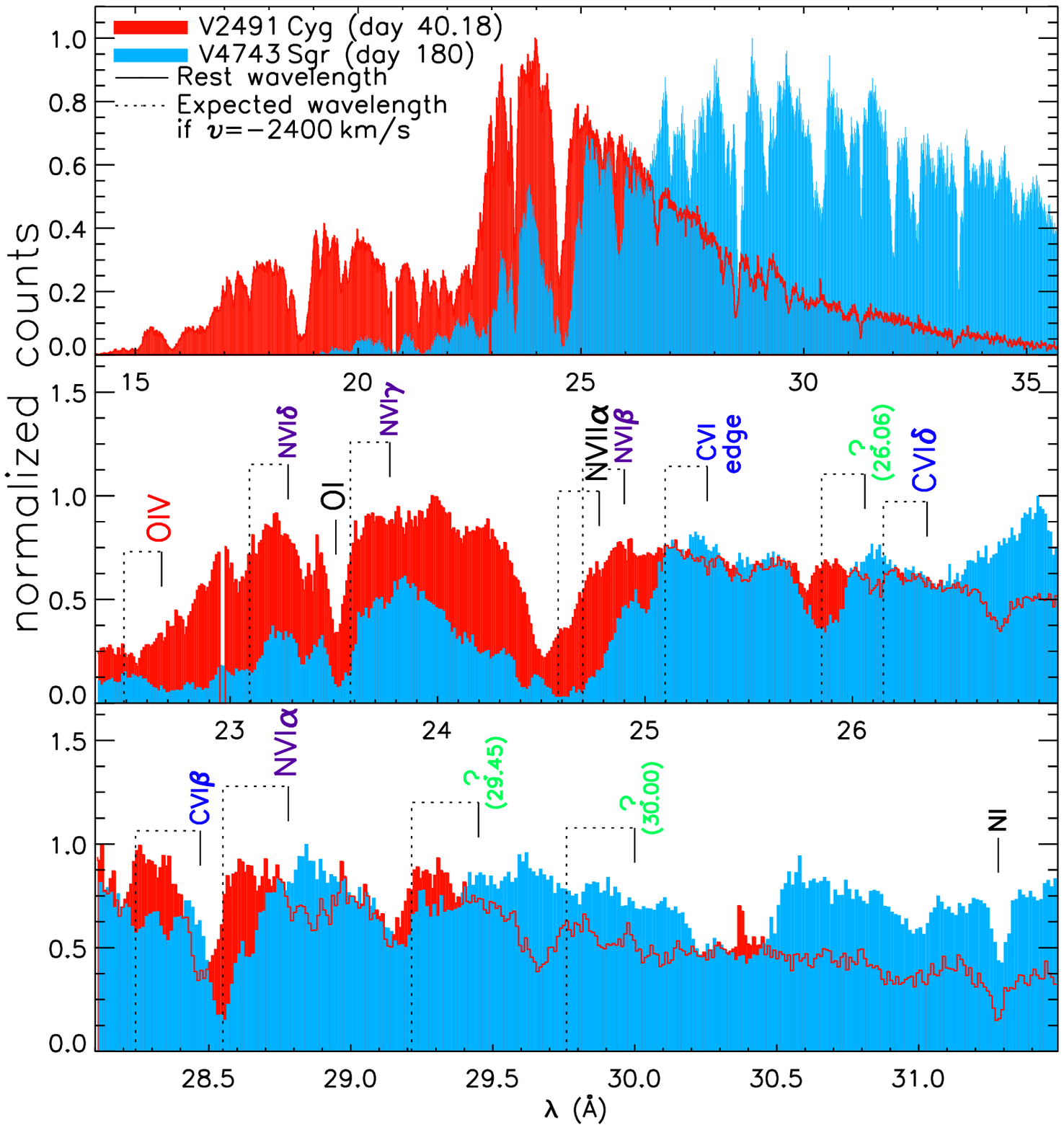}}
\caption{\label{cmp_v4743}Comparison of a calibrated
\chandra/LETGS spectrum of V4743\,Sgr (day 180 after discovery;
light blue) to the phase c spectrum of V2491\,Cyg (starting
day 40.18, red as in Fig.~\ref{lc}). The same lines as in
Fig.~\ref{levol} are marked, and the dotted lines indicate
the expected blue shifts for an expansion velocity of
$-2400$\,km\,s$^{-1}$ \citep{v4743}.
}
\end{figure}

\begin{figure}
\resizebox{\hsize}{!}{\includegraphics{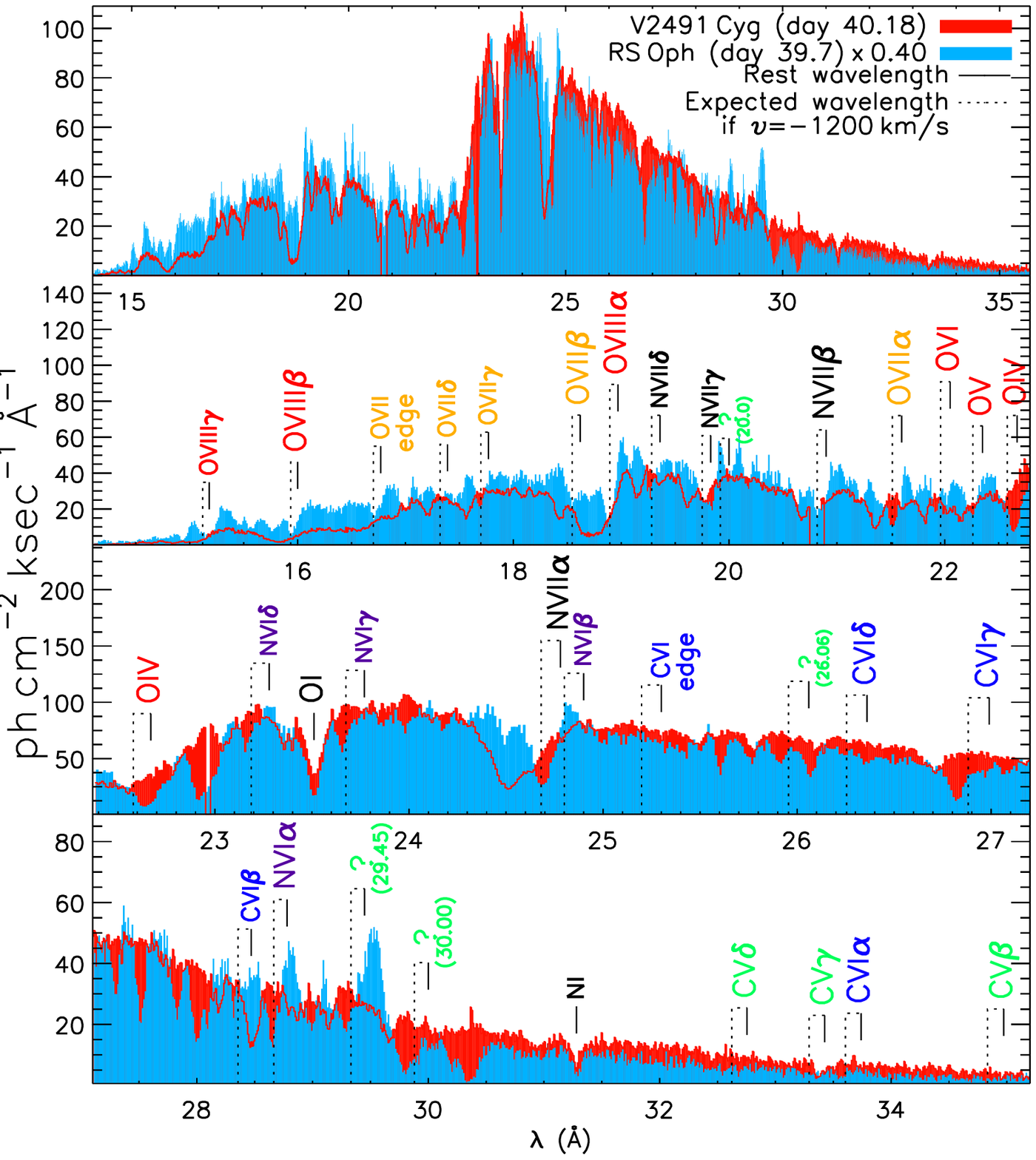}}
\caption{\label{cmp_rsoph}Same as in Fig.~\ref{cmp_v4743}
for RS\,Oph (light blue). Here, line shifts of
$-1200$\,km\,s$^{-1}$ are marked \citep{ness_rsoph}.
}
\end{figure}

\section{Discussion}
\label{disc}

\subsection{Light Curves}

 Based on the second half of the first observation, a periodic
signal of 2232.1 seconds (37.2 minutes, frequency 0.448\,mHz)
may be present. We cannot entirely exclude red noise, but it
is noteworthy that this signal is of the same order
as the $\sim 2500$-s period detected by \cite{drake03} in
V1494\,Aql or the 1300-sec period seen by \cite{v4743} in
V4743\,Sgr. It could be the spin period of the WD or originate
from pulsations \citep{drake03,kang06}. A detailed study of the
X-ray light curves of V4743\,Sgr by \cite{dobrness09} revealed
the presence of two nearby signals between days 180 and 196 (see
also \citealt{leibowitz06}). While
roughly the same period was found to be persistent over the
course of three years, small frequency changes were found
by \cite{dobrness09}.

 During the SSS phase, the frequency changes
scaled with changes in the blackbody temperature as predicted by
stellar pulsation theory (for details see \citealt{dobrness09}).
After day 526, a stable frequency was found in post-outburst
X-ray light curves. However, in optical and UV light curves of
V4743\,Sgr a slightly different frequency was found. They
concluded that while pulsations may have been present during
the SSS phase, the post-outburst X-ray period could be the WD spin
period, modulated by pulsations during the SSS phase.
Combined with the orbital period, \cite{kang06} and \cite{dobrness09}
conclude that V4743\,Sgr is an intermediate polar. For
V2491\,Cyg, only two light
curves are available, the first of which is interrupted by a
dip that has to be excluded from the period analysis.
In intermediate polars, the orbital period is usually
significantly longer than the spin period of the WD, i.e., there
is no spin/orbit synchronization.
\cite{atel1514} found a period of 2.3 hours for V2491\,Cyg,
which is much longer than the 37.2-minute period, but it was
not claimed to be the orbital period. The fact that a period can
be seen in the X-ray light curve could imply that magnetic fields
are involved, but not enough information is available to make a
strong enough case for an intermediate polar on these grounds.
Particularly puzzling is that the period is not present in the
second observation, while the X-ray light curve of an 
intermediate polar should be modulated with the spin
period of the WD. Spectroscopic evidence supporting
the interpretation of an intermediate polar is discussed by
\cite{dai_v2491}.

\subsection{Spectra}

\subsubsection{Applicability of theoretical stellar atmospheres}
\label{atmdisc}

 The ultimate goal of spectral analyses is to determine
mass loss, composition, and
kinematics. The requirements for finding these parameters are
well-exposed X-ray grating spectra that allow individual
lines to be resolved and sophisticated spectral models that
include the requisite physics and reproduce the observed spectra.

X-ray spectra of CCD resolution ($\sim 100$\,eV) are not sufficient,
as spectral models that are different in their details will be
indistinguishable after folding them through the instrumental
response of a CCD spectrometer. Too many different models can
reproduce the same data, yielding non-unique results.
Fine tuning of some parameters may improve
the value of $\chi^2$, but with the limited amount of
information in the observed spectrum, the adjustments in the
model are arbitrary.

 With the \xmm\ and \chandra\ gratings,
appropriate spectra can be taken and are presented in this
paper. The first requirement, well-exposed high-resolution spectra,
has been met, however, spectral models need more development.
The kinematics in the ejecta are relatively easy to determine from
the blue shifts in the absorption lines, but at the same time they
pose the biggest challenge to self-consistent non-LTE atmosphere models.

 Another problem is that we depend on a very limited number of
publicly available models. An attempt has been made by
\cite{atmtables} to make available an archive of TMAP models, covering
a multi dimensional space of parameters. While the approach yields
the most sophisticated models using a large database of transitions
and treating radiation transport in full non-LTE, they are
plane-parallel plus static and as such inappropriate for nova
ejecta while they are still expanding.
We have shown in \S\ref{lprofil}, where Table~\ref{linetab} is
discussed and Fig.~\ref{n7} is shown as an example, that all
photospheric absorption lines are blue shifted as well as
significantly broader than the
instrumental line broadening function. This can only be interpreted
as expanding ejecta that are sufficiently extended to observe a
range of layers of different velocity and consequently also temperature.
\cite{ness09} has shown that this phenomenon can also be observed for
other novae. The TMAP models account for non-LTE effects,
however, the underlying assumptions of hydrostatic equilibrium
and plane parallel geometry are far from reality for
expanding nova ejecta, and our results show that such
simplifications are not justified. For more detailed discussion
of the importance of the missing physics, we refer to 
\cite{hauschildt92,hauschildt94a,hauschildt94b,hauschildt95,phx_expand}.
Although, these arguments apply to an earlier phase of the nova
evolution, the high-resolution spectra indicate that
the arguments are the same. No models are publicly available
that account for all the necessary physics, and we refer to
\cite{vanrossumness09} who found that atmosphere models
that account for the expansion yield vastly different results
from static models.

 Clearly, atmosphere models are required that can give a full
account of the physics in the ejecta without neglecting the
extended nature and/or expansion, including the effects of
spherical and non-spherical geometry. Without such
efforts, no accurate determination of a photospheric temperature
is possible, even if the spectra can be well fitted.

 Meanwhile, blackbody fits are at most parameterizations yielding
no more than a spectral hardness parameter. Some limited, qualitative
studies of spectral changes can be carried out, but no quantitative
interpretation is possible. The advantage of blackbody fits is
the small parameter space, yielding unique parameters, and
blackbody temperature values respond sensitively to changes
in spectral hardness. While different atmospheric processes
may balance out to result in similar temperatures with the
brightness decrease on day 40.03, it appears more likely that only the
brightness has changed while the atmospheric structure remained
the same. This would argue in favor of external reasons for
the brightness variations. Meanwhile, the increase in blackbody
temperature towards the second observation could be explained
by either a higher photospheric temperature or changes in the
atmosphere structure. If interpreted as an increase in photospheric
temperature, this result would be consistent with a decreasing
photospheric radius, exposing successively hotter material.\\

\subsubsection{Comparison to other novae}
\label{cmpothernovae}

 V2491\,Cyg is the third nova for which a bright SSS X-ray
spectrum with atmospheric continuum and deep absorption lines
was obtained in high spectral resolution, after V4743\,Sgr and
RS\,Oph, and a comparison is instructive.

{\bf V4743\,Sgr}\\
In Fig.~\ref{cmp_v4743}, a comparison between a calibrated
\chandra/LETGS observation of the slower ($t_2=9$ days,
\citealt{v4743_t2}) nova V4743\,Sgr as
observed on day 180 after discovery (light blue) and the
day 40.18 spectrum of V2491\,Cyg (red, see Table~\ref{sepoch})
is shown.
 The spectrum of V4743\,Sgr is significantly softer than
V2491\,Cyg,
indicating a lower photospheric temperature. While the lower
$E(B-V) = 0.25$ for V4743\,Sgr (as opposed to $E(B-V)=0.43$
for V2491\,Cyg by \citealt{rudy08}) could explain a softer X-ray
spectrum, the Wien tail being at longer wavelengths makes a
strong case for a lower effective temperature. In the bottom
panels, the two wavelength ranges 22-27\,\AA\ and 27-31\,\AA\
are shown in
detail. The same line labels as those used in Fig.~\ref{levol} are
included, but here, we assume a blue shift corresponding to
$-2400$\,km\,s$^{-1}$, which is consistent with several
photospheric lines in V4743\,Sgr such as N\,{\sc vii} at 25\,\AA\
\citep{v4743}. The line profiles in V4743\,Sgr are much more
complex than in V2491\,Cyg, some of them consisting of more than
one velocity component. A similarly complex situation is encountered
in KT\,Eri \citep{ATel2418}. For example, the line
profile around the N\,{\sc vii} line is extremely complex in
V4743\,Sgr, while
for V2491\,Cyg, only two overlapping lines from N\,{\sc vii} and
N\,{\sc vi} are seen (see Fig.~\ref{n7}).
The interstellar N\,{\sc i} and O\,{\sc i}
lines at 32.28\,\AA\ and 23.5\,\AA, respectively, are present in
both novae at the same wavelengths. The O\,{\sc i} line
blends with the blue-shifted N\,{\sc vi}$\gamma$ line in V4743\,Sgr.

{\bf RS\,Oph}\\
More similar to V2491\,Cyg is a grating spectrum taken 39.7 days
after the discovery of the 2006 outburst of the recurrent, symbiotic
nova RS\,Oph \citep{ness_rsoph} which evolved
on a similar time scale ($t_2=7.9$ days, \citealt{hounsell10}).
This is shown in Fig.~\ref{cmp_rsoph} where
the calibrated \chandra/LETGS spectrum of RS\,Oph (in flux units)
was scaled by a factor 0.4. The observed blue shifts seen in
the photospheric
lines in RS\,Oph was $-1200$\,km\,s$^{-1}$ \citep{ness_rsoph}, and the
dotted lines mark these projected velocities.

The scaling factor of 0.4 is much larger than is required
for the different distances of 1.6\,kpc for
RS\,Oph \citep{bods87} and 10.5\,kpc for V2491\,Cyg. Thus, at
the respective times of observation, V2491\,Cyg was intrinsically
a factor 17 brighter than RS\,Oph. However, both novae were extremely
variable around the respective times of observations.

 The spectrum
of V2491\,Cyg used for the comparison has been extracted from a
phase of bright emission (day 40.18) while the \chandra\
observation of RS\,Oph was taken during a time of comparatively
low flux \citep{ness_rsoph}. In addition, there are uncertainties
in the distance of V2491\,Cyg, because the temporary secondary
peak seen in V2491\,Cyg (see bottom panel of Fig.~\ref{swlc})
disallows the use of the MMRD method.
Under the assumption of the same luminosity at the respective
times of observation, the scaling factor between RS\,Oph and
V2491\,Cyg yields a distance of 2.5\,kpc for V2491\,Cyg.
The soft tails of the X-ray spectra of V2491\,Cyg and RS\,Oph
are very similar, indicating the same amount of photoelectric
absorption, which is consistent with the measurements of
$E(B-V)=0.43$ for V2491 Cyg (\citealt{rudy08}, corresponding
to $N_{\rm H}=2.6\times10^{21}$\,cm$^{-2}$) and
$N_{\rm H}=2.4\times10^{21}$\,cm$^{-2}$ for RS\,Oph
\citep{hje86}.

Noteworthy is a remarkable similarity in the continua between the
X-ray grating spectra of V2491\,Cyg and RS\,Oph (Fig.~\ref{cmp_rsoph})
suggesting a surprisingly similar photospheric temperature at the times
of the respective observations. With CCD resolution, V2491\,Cyg and
RS\,Oph would be considered twins.
However, at grating
resultion, small but important differences can be seen, yielding
information about the structure of the atmosphere.
Most obvious are much smaller blue shifts and line widths of the
photospheric lines in RS\,Oph, indicating slower expansion.
Also, the O\,{\sc i} absorption edge at 23.5\,\AA\ is
slightly steeper in V2491\,Cyg. While the N\,{\sc vi} and
N\,{\sc vii} lines around 28.7 and 25\,\AA\ reach about the
same depth in both cases,
the O\,{\sc viii} line around 19\,\AA\ is much deeper in V2491\,Cyg.
Since the same depth is reached for the lines of two different
ionization stages, N\,{\sc vi} and N\,{\sc vii}, the deeper
O\,{\sc viii} line cannot be a temperature effect, and a higher
oxygen abundance in V2491\,Cyg appears likely. This could also
explain the deeper O\,{\sc i} absorption edge. On the other hand,
the O\,{\sc vii} line at 21.6\,\AA\ is only slightly deeper in
V2491\,Cyg, but that could be explained by photoionization from
O\,{\sc vii} to O\,{\sc viii}, which is supported by the
presence of the O\,{\sc vii} absorption edge at 16.7\,\AA\ in
V2491\,Cyg (see Fig.~\ref{o7}) that seems less pronounced in RS\,Oph.

{\bf Unidentified lines}\\
While the optical evolution was different, yielding
a rebrightening in V2491\,Cyg that was not seen in RS\,Oph,
V2491\,Cyg was suggested to be a recurrent nova
\citep{v2487oph_rn,page09,v2491cyg_rn}, and the spectral similarity to
one of the most famous recurrent novae, RS\,Oph, may be an additional
indicator.

 While the measurement of line shifts relies on well-identified lines,
a number of absorption lines cannot be identified.
 The comparison with other novae gives us an idea of the formation
conditions of unidentified lines, aiding future identification
efforts.
At 25.85\,\AA, a broad line is seen in V4743\,Sgr that coincides
with a line at 26.06\,\AA\ that is marked as unidentified in
Fig.~\ref{levol}. With the lower photospheric temperature of
V4743\,Sgr, this line could be formed at relatively lower
temperature since it is much more pronounced in V4743\,Sgr.
Another unidentified line in Fig.~\ref{levol}, with a
projected rest wavelength of 29.45\,\AA, is also present
in V4743\,Sgr with roughly the same width and strength. The
only difference is the line shift, which is similar
to the N\,{\sc vi} line at 28.8\,\AA. This line might be formed
at a slightly higher temperature than N\,{\sc vi} which is deeper
in V4743\,Sgr. The line at 30\,\AA\ projected rest wavelength
is not present in V4743\,Sgr and could be formed at
temperatures that are higher than in V4743\,Sgr.
The unidentified lines are listed in Table~\ref{unidentified}
with projected rest wavelengths in the first column and
observed wavelengths for V2491\,Cyg, V4743\,Sgr, and
RS\,Oph (see below), respectively.

 The unidentified lines with projected rest wavelengths 26.06,
29.45, and 30.0\,\AA\ are also present in RS\,Oph. The 26.06-\AA\ line
is weaker in RS\,Oph than in V4743\,Sgr, but stronger than in
V2491\,Cyg, while the 29.45-\AA\ line has about the same strength.
This supports the conclusion that the 26.06-\AA\ line is
a low-temperature line while the 29.45-\AA\ line is formed at
higher temperatures. In contrast to V4743\,Sgr, the 30.0-\AA\
line is clearly present in RS\,Oph and is deeper than in
V2491\,Cyg. Additional unidentified lines are seen
in RS\,Oph, e.g., at 25.6, 25.7, 26.1, 27.3, 27.5, 27.6, 28.0,
and 30.3\,\AA, but these lines are not seen in V2491\,Cyg and
V4743\,Sgr.
Two of the unidentified lines are seen in all three novae at
wavelengths yielding the same blue shifts as other photospheric
lines. Meanwhile, a large number of lines are only present in the
spectrum of RS\,Oph, arising in the same wavelength range as
a pre-SSS emission line component around day 26
\citep{rsophshock,nelson07}.

The emission lines in the RS\,Oph spectrum at 28.8, 29.1,
and 29.5\,\AA\ originate from the N\,{\sc vi} He-like triplet.
These and other
components of the red wings of absorption lines such as
O\,{\sc viii} (18.97\,\AA) and N\,{\sc vii} (24.8\,\AA)
are related to the P Cygni-like profiles discussed by
\cite{ness_rsoph}. They could partially originate in shocked
plasma and might not be related to the atmospheric emission.

{\bf Other novae}\\
Other nova SSS spectra such as those of
V1494\,Aql \citep{rohrbach09}, are more complicated and/or less
well exposed. All high-quality SSS spectra show significant blue shifts in
their absorption lines, and the conclusion is unavoidable
that expansion continues into the SSS phase \citep{ness09}.
It is noteworthy that the expansion velocities can be quite
different, yielding the lowest expansion velocity in RS\,Oph,
even though the spectral shape of the continuum and the depths
of most absorption lines are remarkably similar to V2491\,Cyg.
The lower expansion velocity in RS\,Oph may be explained by the
presence of a dense stellar wind of the companion that has absorbed
some of the kinetic energy during the earlier phases of the outburst.
This results in significant deceleration occurring by $\sim 40$ days
after outburst \cite[e.g.][]{bode06}.
While some evidence for the companion in V2491\,Cyg to be an evolved
object is presented by \cite{ribeiro10},
the mass loss rate is likely to be much lower than in RS\,Oph.
Meanwhile, in V4743\,Sgr, some of the
absorption lines contain at least three velocity components
while V2491\,Cyg and RS\,Oph contain only one velocity
component. Since V4743\,Sgr is a slower nova, it may be possible
that the ejecta in slower novae are generally more structured
than those of faster novae. This would be consistent with modeling
of common envelope interaction \citep{lloyd95a,porter98}, but
other slower novae need to be studied for firmer conclusions.\\

\subsubsection{High-amplitude variations}

\begin{figure}
\resizebox{\hsize}{!}{\includegraphics{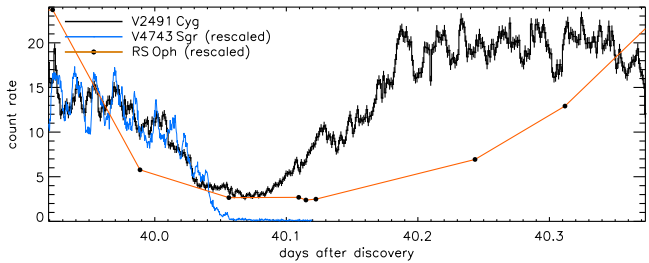}}
\caption{\label{cmplc_v4743}Comparison of the X-ray light curve of
V2491\,Cyg after day 39.6 with rescaled \chandra\ and \swift\ light
curves of V4743\,Sgr with the decay on day 180.6 and of RS\,Oph with
one of the minima reported by \cite{osborne11}, respectively.
}
\end{figure}

 During the earlier of the two \xmm\ observations, lasting from day
39.93 to 40.37, V2491\,Cyg was highly variable in X-rays, including a
three-hour dip. Several novae have been observed with an episode of
high-amplitude variations during their early SSS phase, but an
explanation is still pending. The first observed occurrence of a
sudden decay in the X-ray output of a nova was encountered in a
\chandra\ observation of V4743\,Sgr (see Fig.~\ref{cmplc_v4743} and
\citealt{v4743}). The count rate declined almost to zero, leaving behind
an emission line spectrum, originating from photoionization plasma with
some emission lines that were rapidly declining (Ness et al. in prep).
At that time, no monitoring
was possible, and the nova did not rebrighten during the same
observation. Two weeks later, the nova was bright again, but highly
variable \citep{leibowitz06}. The dip in the first \xmm\ observation
of V2491\,Cyg occurred on the same time scale as the decay in
V4743\,Sgr, as illustrated in Fig.~\ref{cmplc_v4743}, and the
same phenomenon may have been observed.

Dense \swift\ monitoring of RS\,Oph revealed multiple declines
\citep{osborne11} which occurred on time scales that are also similar
to V2491\,Cyg (Fig.~\ref{cmplc_v4743}). This phenomenon has not been
fully explained yet. One suggestion is temporary photospheric expansion,
which would be supported by the disappearance of periodic oscillations
during the dip which likely originate from close to the WD.
Such deep dips could also
mean that the concept of constant bolometric luminosity is
wrong as suggested by \cite{10chan}. However,
intrinsic changes in bolometric luminosity are not necessarily
easier to understand than changes in the structure of
the ejecta.
A change in luminosity would imply a change in the ionization
parameter, $\xi=L/(nr^2)$, of the photo-ionized gas, with a
subsequent evolution of the photo-ionization balance. Since
the density of the photo-ionized absorbing gas, $n$, and distance to
ionizing source, $R$, are unlikely to change significantly,
intrinsic changes of the bolometric luminosity would have lead to
equally significant changes in the ionization parameter. A detailed
study using photo-ionization models is in progress
(Pinto et al. in prep).

\section{Summary and Conclusions}
\label{summary}

 A small sample of extremely well exposed grating spectra of novae
during their SSS phase has been gathered. The fast classical nova
V2491\,Cyg was observed twice during its extremely short SSS phase,
covering two different phases of evolution. The first observation
was taken early while the X-ray brightness was highly variable,
containing a deep dip, while the second observation was taken
during the decline with a lower count rate and less variability.
The dip evolved on a time scale similar to the decay in the
slower nova V4743\,Sgr \citep{v4743} and the variations during
the early SSS phase in RS\,Oph \citep{osborne11}. High-amplitude
variations seem to be a common phenomenon that has now been
observed in many novae.

The contemporaneous UV light curves show short-term variability,
but no UV data were taken during the dip in the first X-ray light curve.
A 37.2-minute period was detected in X-rays before and after but not
during the dip. It could reflect the WD spin period, but the
statistical significance is not high.

The large majority of absorption lines in the high-resolution
RGS X-ray spectra yield a single velocity component with blue shifts
ranging between $-3000$ and $-3400$\,km\,s$^{-1}$.
Additional interstellar lines are found at their rest wavelengths.
A few deep absorption lines are currently not identified.
The same lines are also present in the SSS spectra of V4743\,Sgr
and RS\,Oph at a different blue shift, indicating that they are
of photospheric origin.

 The narrow range of blue shifts is
different from V4743\,Sgr where at least three velocity components
can be seen. This could be explained by the slower evolution of
V4743\,Sgr. The SSS spectrum of RS\,Oph is remarkably similar in
the shape of the continuum, yet is very different in the properties
of the absorption lines, yielding much smaller blue shifts.
The lower velocities in RS\,Oph may be due to the presence of the
pre-outburst red giant wind, while the higher blue shifts in
V2491\,Cyg could be the explanation for the short duration of
the SSS phase.

 Archived atmosphere models yield good fits to the SSS spectra,
but the model parameters are unphysical. This can be explained
by the simplifications in the model, assuming hydrostatic equilibrium
and a compact atmosphere without expansion.\\

The conclusions are the following:
\begin{enumerate}
\item Significant blue shifts in all photospheric absorption lines
demonstrate that the nova ejecta are expanding.
\item Significant line broadening of most photospheric absorption
lines demonstrates that the ejecta are extended, exposing a range of
velocity layers and thus also temperature layers.
\item Our conclusions from line profile analysis indicate that
publicly available stellar atmosphere models do not contain enough
physics for nova ejecta,
even though they yield surprisingly good fits to the data.
\item Comparison with other novae of lower and higher photospheric
temperatures allows us to spot so far unknown nova absorption lines
and associate them to higher or lower temperature environments.
\item Remarkable spectral similarities between V2491\,Cyg and RS\,Oph
indicate similar system characteristics. Assuming the same luminosities,
the distance to V2491\,Cyg would only be 2.5\,kpc.
\item The different properties of the absorption lines between
V2491\,Cyg and RS\,Oph indicate that similar global parameters
can still lead to a different atmospheric structure.
\item Slower novae such as V4743\,Sgr may contain more complex
absorption line systems than faster novae such as RS\,Oph or
V2491\,Cyg.
\item The single dip in the X-ray light curve occurred on the
same time scale as in V4743\,Sgr and RS\,Oph, and the underlying
process may therefore be common to all novae in which an early
variation phase is observed in X-rays.
\end{enumerate}

\acknowledgments

We thank the \xmm\ SOC team for their support in scheduling and analyzing
these TOO observations. Special thanks to Aitor Ibarra who supported us
with the new SAS tool {\tt xmmextractor}.
We acknowledge with thanks the variable star observations from the AAVSO International Database contributed by observers worldwide and used in this research.
J-UN gratefully acknowledges support provided by NASA
through \chandra\ Postdoctoral Fellowship grant PF5-60039,
awarded by the \chandra\ X-ray Center, which
is operated by the Smithsonian Astrophysical Observatory for NASA
under contract NAS8-03060. A.D. and C.P. acknowledge financial support
from the Faculty of the European Space Astronomy Centre.
AD was supported in part by the Grand-in-Aid for the global COE programs on "The Next Generation of Physics, spun from Diversity and Emergence" from MEXT and also by the Slovak Grant Agency, grant VEGA-1/0520/10.
KLP, APB, and JPO acknowledge support from STFC.
JJD was  supported by NASA contract
NAS8-39073 to the {\em Chandra X-ray Center} during the course of this
research.
SRON Netherlands is supported financially by NWO, The Netherlands
Organization for Scientific Research.
SS acknowledges support from NASA+NSF grants to ASU.
We thank Rick Rudy and Dave Lynch for the information on the peak
at the later NIR spectra.
MH acknowledges support from MICINN project AYA 2008-01839/ESP,
AGAUR project 2009 SGR 315 and FEDER funds and GS MICINN projects AYA
2008-04211-C02-01 and AYA 2007-66256.

\bibliographystyle{apj}
\bibliography{cn,astron,jn,rsoph}

\end{document}